\documentclass[letterpaper, 10 pt, journal]{ieeeconf}  % Comment this line out if you need a4paper

\IEEEoverridecommandlockouts                              % This command is only needed if 
\renewcommand{\baselinestretch}{0.955}

\usepackage{amsmath}
\usepackage{amssymb}
\usepackage{tikz}

\usepackage{amsthm}
\usepackage{xspace}
\usepackage{array}
\usepackage{multirow}

\usepackage{colortbl}
\definecolor{cello}{HTML}{ffe6cc}

\usepackage{graphics} % for pdf, bitmapped graphics files

\newtheorem{mytheorem}{Theorem}

\usepackage{color}
\usepackage{xcolor}
\usepackage{accents}
\usepackage{centernot}
\usepackage{algorithm}
\usepackage{algorithmic}
\usepackage[font=small, labelfont=bf]{caption}
\usepackage{mathtools}

\def \Vh0{\stackrel{\circ}{V}_h}

    \def\R{{\mathbb R}}

   \def\eps{\varepsilon}

\newcommand{\lc}
{\mathrel{\raise2pt\hbox{${\mathop<\limits_{\raise1pt\hbox
{\mbox{$\sim$}}}}$}}}

\newcommand{\gc}
{\mathrel{\raise2pt\hbox{${\mathop>\limits_{\raise1pt\hbox{\mbox{$\sim$}}}}$}}}

\newcommand{\ec}
{\mathrel{\raise2pt\hbox{${\mathop=\limits_{\raise1pt\hbox{\mbox{$\sim$}}}}$}}}

\def\bb{\begin{equation}} \def\ee{\end{equation}}
\def\beqn{\begin{eqnarray}}  \def\eqn{\end{eqnarray}}
\def\beqnx{\begin{eqnarray*}} \def\eqnx{\end{eqnarray*}}
\def\bn{\begin{enumerate}} \def\en{\end{enumerate}}

\def\bd{\begin{description}} \def\ed{\end{description}}
\makeatletter
\let\NAT@parse\undefined
\makeatother
\usepackage{hyperref}

\title{\LARGE \bf
Koopman-Hopf Hamilton-Jacobi Reachability and Control % tempting
}

\author{Will Sharpless, Nikhil U. Shinde, Matthew Kim, Yat Tin Chow and Sylvia Herbert
\thanks{Sharpless, Shinde, Kim, and Herbert are with University of California, San Diego. Chow is with the University of California, Riverside. 
\{\href{mailto:wsharpless@ucsd.edu}{wsharpless}, \href{mailto:nshinde@ucsd.edu}{nshinde}, \href{mailto:mak009@ucsd.edu}{mak009}, \href{mailto:sherbert@ucsd.edu}{sherbert}\}@ucsd.edu, \href{mailto:yattinc@ucr.edu}{yattinc}@ucr.edu
This work is supported by NIH training grant T32 EB009380 (McCulloch), NSF DMS-2409903, ONR N000142412661 and ONR YIP N00014-22-1-2292. The content is solely the responsibility of the authors.}
}

\begin{document}
\maketitle
% \captionsetup[algorithm]{labelformat=empty}
% \algrenewcommand\algorithmicrequire{\textbf{Input:}}
\thispagestyle{empty}
\pagestyle{empty}
\begin{abstract}
The Hopf formula for Hamilton-Jacobi reachability (HJR) analysis has been proposed to solve high-dimensional differential games, producing the set of initial states and corresponding controller required to reach (or avoid) a target despite bounded disturbances. 
As a space-parallelizable method, the Hopf formula avoids the \textit{curse of dimensionality} that afflicts standard dynamic-programming HJR, but is restricted to linear time-varying systems.
To compute reachable sets for high-dimensional nonlinear systems, we pair the Hopf solution with Koopman theory for global linearization. 
By first lifting a nonlinear system to a linear space and then solving the Hopf formula, approximate reachable sets can be efficiently computed that are much more accurate than local linearizations. 
Furthermore, we construct a Koopman-Hopf disturbance-rejecting controller, and test its ability to drive a 10-dimensional nonlinear glycolysis model. We find that it significantly out-competes expectation-minimizing and game-theoretic model predictive controllers with the same Koopman linearization in the presence of bounded stochastic disturbance. In summary, we demonstrate a dimension-robust method to approximately solve HJR, allowing novel application to analyze and control high-dimensional, nonlinear systems with disturbance. An open-source toolbox in Julia is introduced for both Hopf and Koopman-Hopf reachability and control.

%Standard dynamic-programming based approaches for computing the BRS are computationally intractable for high-dimensional (6D+) systems. The recently introduced Hopf formulation for reachability anaylsis solves HJ reachability problems using an efficient space-parallelizable method. In exchange, however, a complex optimization problem must be solved, limited in application to linear time-varying systems. 

%has been proposed for Hamilton-Jacobi Reachability (HJR) for solving high-dimensional differential games as a space-parallelizeable method. In exchange, however, a complex optimization problem must be solved, limited in application to linear time-varying systems. To compute the HJ backwards reachable set (BRS) for high-dimensional, nonlinear systems, we pair the Hopf solution with Koopman theory for global linearization. We find that this is a viable method for approximating the BRS and performs significantly better than a local linearization in the Slow Manifold and Duffing systems. Furthermore, we construct a Koopman-Hopf controller based on the HJR disturbance-rejecting controller and test it's ability to drive a 10-dimensional glycolysis model. We find that it significantly out-competes expectation-minimizing and game-theoretic model predictive controllers with the same Koopman linearization in the presence of bounded stochastic disturbance. In summary, we demonstrate a dimension-robust method to approximately solve HJR, allowing novel application to analyze and control high-dimensional, nonlinear systems with disturbance.
\end{abstract}

\section{Introduction}

There are several directions for designing safe controllers for autonomy. The technical rigor involved in planning for success while avoiding failure tends to force methods to sacrifice guarantees (e.g., data-driven methods) or feasibility due to over-conservative solutions (e.g. differential inclusions).  

Among these approaches, Hamilton-Jacobi reachability (HJR) is well known for being a robust approach to optimal control and safe path planning \cite{bansal2017hamilton, chen2017exact, donggun19iterativehopf, Kirchner_2018, chow2017algorithm}. This method is usually used to generate the backward reachable set (BRS) of a system: the set of states from which a system with bounded control can reach (or avoid) a target  despite bounded disturbances. This analysis also generates a corresponding optimal controller that rejects bounded disturbances. When feasible, it is a powerful tool for autonomous guidance and other stochastic control problems because of its derivation from the theory of differential games \cite{evans1984differential, lions1986hopf} which describes how to optimally drive a system to counter antagonistic or stochastic, bounded disturbances \cite{bacsar1998dynamic}. HJR is, however, often not the most practical approach because of its dependency on spatial gradient approximations in a dynamic-programming (DP) scheme which makes it sensitive to the \textit{curse of dimensionality} \cite{bansal2017hamilton}. If this theory could be extended to higher dimensional systems, engineering efforts in diverse domains, particularly in medicine, finance and other large systems, could make strides where simpler (dimension-robust) controllers are unable to overcome disturbances or model uncertainty. 

\begin{figure}[t]
    \centering
    \includegraphics[width=0.9\linewidth]{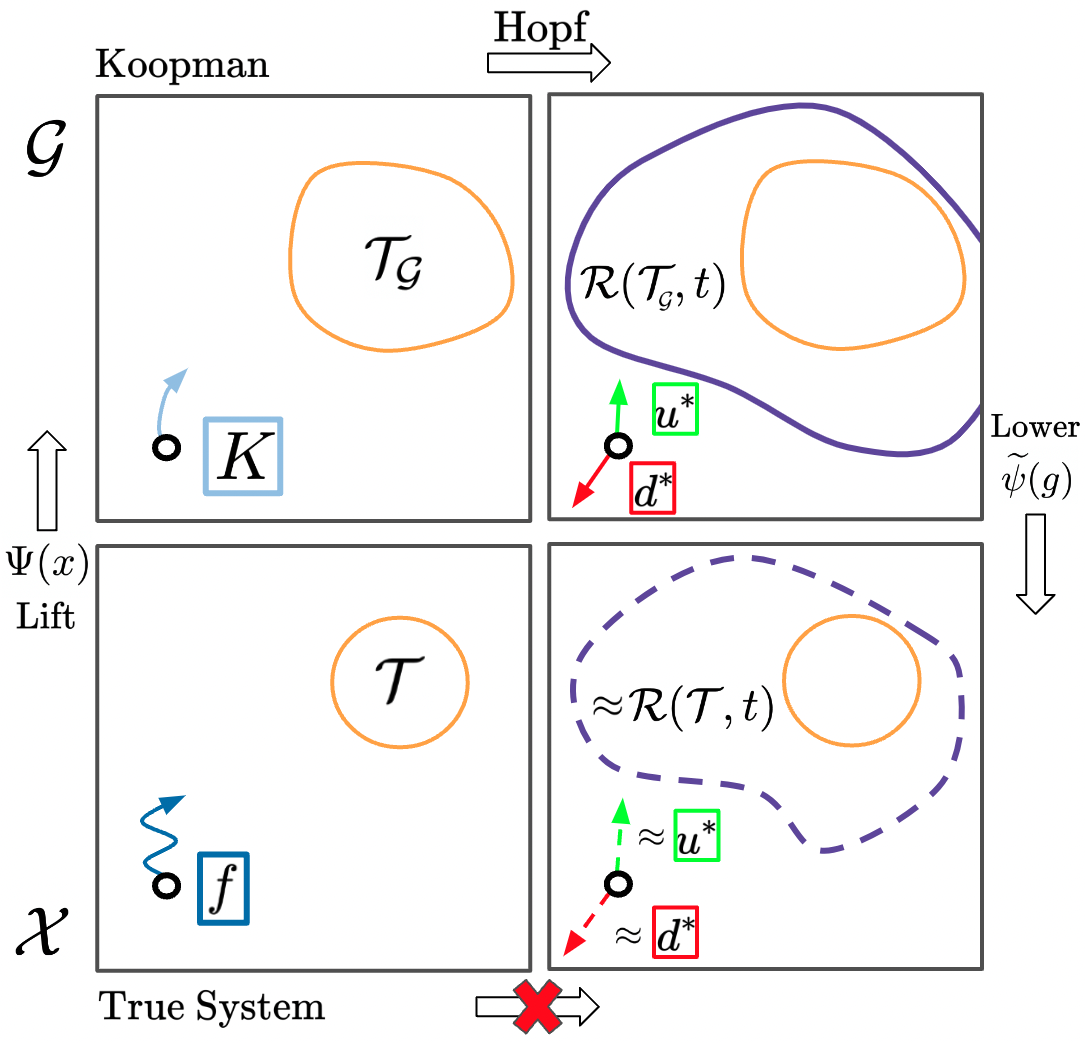}    
    \caption{\textbf{Graphical Overview} The bottom left panel depicts the problem formulation, wherein a system with nonlinear dynamics seeks to either reach or avoid a target set $\mathcal{T}$. The system may have bounded control and disturbances, formulating a two-player game. The upper-left panel shows the same problem formulation in the lifted linear space, where via Koopman theory the same system is represented by a high-dimensional linear system seeking to reach or avoid the ``lifted target'' $\mathcal{T}_\mathcal{G}$. The upper-right panel depicts the use of the Hopf solution in this setting to solve the value of the lifted target, yielding the lifted reachable set $\mathcal{R}(\mathcal{T}_\mathcal{G}, t)$ and optimal control $u^*$ and disturbance $d^*$. Finally, the lower-right panel depicts the mapping of the solution to the original space to approximate the true reachable set $\mathcal{R}(\mathcal{T}, t)$ and optimal control and disturbance. 
    \vspace{-2em}}
    \label{fig:Abstract_Graphic} 
\end{figure}

% \begin{figure}[t]
%     \centering
%     \includegraphics[width=\linewidth]{BRS_duffing_2s_scatter_nice.png}    
%     \caption{Here the $\pm \epsilon$-boundary of a target $\mathcal{T}$ (black), the DP-solved reachable set $\mathcal{R}(\mathcal{T},t)$ (blue), the local-linearization Hopf BRS $\mathcal{R}( \mathcal{T}_{\text{Taylor}},t)$ (green), and our 15D Koopman-Hopf BRS $\mathcal{R}(\widetilde{\mathcal{T}}_\mathcal{G},t)$ (gold) are plotted for the Duffing oscillator at $t=2s$ with control and disturbance and $\epsilon=0.1$.  We show the scatter plot to emphasize that Koopman-Hopf solution is solved in a space-parallizeable fashion, hence, each point (i.e. ($t$,$x$)) is solved independently.
%  %\zgnote{for this figure, seems like the gold one approximates the best right? Maybe try to use bigger points to emphisize the good results and use smaller points to weak the 'not so good' resutls. (or maybe darker vs lighter color)}
%     \vspace{-2em}}
%     \label{fig:BRSapprox} 
% \end{figure}

\subsection{Related Work}

Toward this end, several directions have been developed, including set-based propagation, such as the method of zonotopes \cite{althoff2011zonotope}, to over-approximate the Hamilton-Jacobi reachable sets with linear systems \cite{althoff2011zonotope,Bak} and in some special classes of nonlinear systems \cite{majumdar2014convex}. The only shortcomings with these methods are that they do not inherently provide an optimal controller and also tend to be overly-conservative.

Another direction is the method of decomposition and system reduction for HJR. The authors of \cite{chen2017exact} define the system structures that can be decomposed into exponentially-faster, low-dimension problems, however, any coupled dimensions cannot be decomposed. The extension to general systems \cite{chen2018decomposition} tends to be overly conservative. There is also a method of projecting coupled systems to independent, lower-dimensional subsystems such that the inverted solution is a conjectured over-approximation \cite{Ian03}, however, this is often highly conservative and lacks guarantees.

More recently, the Hopf formula was revived \cite{darbon2016algorithms,chow2017algorithm,chow2019algorithm} as an approach to HJR without sensitivity to dimension or sacrificing guarantees. This method involves interchanging the HJR DP problem for an abstract optimization problem over the characteristic curves of the Lagrange multiplier \cite{darbon2016algorithms}. This converts the \textit{curse of dimensionality} issue to a \textit{curse of complexity}, because the optimization problem can be non-convex and non-differentiable and requires sophisticated approaches (see \cite{darbon2016algorithms, chow2017algorithm, chow2019algorithm} for detailed analysis). Nonetheless, certain classes of systems, namely linear time-varying systems, can be robustly solved in both one-player \cite{doggn2020hopf, donggun19iterativehopf} and two-player optimal cases \cite{Kirchner_2018, chow2017algorithm}. Notably, in \cite{Kirchner_2018} Hopf reachability was paired with a naive linearization method to control a pursuit-evasion nonlinear system with success, although without random disturbance. 

We build on the aforementioned strides by novelly pairing the Hopf solution with Koopman theory for global linearization. Koopman theory involves ``lifting'' nonlinear dynamics to high dimensional spaces in order to linearize them with higher accuracy \cite{koopman1931hamiltonian, mezic2021koopman}. Multiple works have found that the Koopman procedure can yield highly accurate predictions over a short horizon, suitable for model predictive control (MPC) and linear quadratic regulators (LQR) \cite{proctor2014dynamic, korda2018linear, yeung2019learning,CompKoop,kaiser2021data}. Furthermore, we are inspired by recent, impressive work pairing Koopman theory with the method of zonotopes for non-optimal reachability \cite{Bak}. 

% Our resulting Koopman-Hopf method can be used to solve HJ backwards reachable sets (BRSs); a BRS is the set of states from which the system can reach a target set at a desired time. A BRS generally may expand in all directions of the state space, and we hypothesized a spatially-balanced linearization would be necessary for accuracy. We find that this has much to do with \cite{Ian03} and solving HJR in a ``parallel" space. Although optimizing the resulting Hopf problem is not simple, we find that with sophisticated formulations such as the alternating direction method of multipliers (ADMM), close approximations of the true BRS can be achieved (see Figure~\ref{fig:BRSapprox}). Moreover, we show that the resulting controller fares better than a stochastic Koopman-MPC (expectation-minimizing) or game-theoretic Koopman-MPC for a 10-dimensional glycolysis model. Ultimately, the accuracy of the BRSs and the robust navigation through a high dimensional, stochastic nonlinear system defend that the proposed pairing is not only valid, but a prudent choice when juxtaposed with other linearization methods or controllers in the Koopman space.

In this work, we use Koopman theory to define a ``lifted'' linear reachability problem which approximates our true problem and can be solved by the dimension-insensitive Hopf solution. Note, this sacrifices guarantees, but yields approximations of BRSs and their corresponding optimal controllers for high-dimensional, nonlinear systems for which the only other method is end-to-end learning \cite{bansal2021deepreach}. A graphical overview of the proposed work can be seen in Figure~\ref{fig:Abstract_Graphic}.

\subsection{Contributions and Organization}

\noindent We make the following contributions:
\begin{enumerate}

    \item We formulate a novel method to approximately solve nonlinear differential games avoiding the \textit{curse of dimensionality}.

    \item We propose definitions for the corresponding Koopman reachability problem that satisfy the Hopf assumptions, and we compare them in the canonical Koopman problem, the slow manifold system.
    
    \item We compare the BRS of our method to that of the DP solution and a Taylor-based solution for both a convex and non-convex game in the Duffing system.
    
    \item We synthesize a novel Koopman-Hopf controller and compare it to two Koopman-MPC formulations in a 10-D glycolysis model with bounded, stochastic disturbance.

    \item We introduce an open-source package \href{https://github.com/UCSD-SASLab/HopfReachability}{\textit{HopfReachability.jl}}, for solving 2-player linear differential games.
    
\end{enumerate}

The paper is structured as follows. Sec. \ref{sec:HJR} formally introduces HJR, BRSs and the DP solution. Sec. \ref{subsec:hopf} introduces the Hopf solution and it's limitations. Sec. \ref{sec:Koop} introduces Koopman theory to counter these limitations. Sec. \ref{sec:koopman_hopf} proposes a method of lifting the reachability problem to the Koopman space that satisfies the Hopf assumptions. Sec. \ref{sec:results} demonstrates the Koopman-Hopf method, first, in the Slow Manifold, and second, in the Duffing system where (because of the low dimensionality) its feasible to compare to DP, and third, in a 10D glycolysis system where we compare the novel controller with Koopman-MPCs. Finally, Sec. \ref{sec:conclusion} summarizes the work and describes future directions.

% Sec. \ref{subsec:BRSapprox} demonstrates the proposed method by comparing the BRS of the true solution in both convex and non-convex games in the Duffing system. Sec. \ref{subsec:Glycolysis} 

% \begin{enumerate}
%     \item HJR
%     \item Linear reachability (zonotopes etc)
%     \item Hopf, 1-Player or naive linearizations 
%     \item Koopman, Koopman-MPC, Koopman-LQR
%     \item Koopman Reachability (stanley bak)
% \end{enumerate}

% \subsection{Hopf Reachability Analysis of 2-Player Games}
% \begin{enumerate}
%     \item great for LTV systems with bounds on input, disturbance
%     \item if nonlinear, optimization of the 2-Player Hopf solution suffers from the curse of complexity
% \end{enumerate}

% \subsection{Koopman Lifting of Nonlinear Systems}
% \begin{enumerate}
%     \item can capture nonlinearities by lifting to higher dim
%     \item since hopf can handle higher dimensions, we propose merging these
% \end{enumerate}

\section{Preliminaries}\label{sec:prelims}
This paper focuses on control-affine and disturbance-affine systems of the form
\beqn
\dot x = f_x(x, t) + h_1(x)u + h_2(x)d \triangleq f(x, u, d, t)
\label{Dynamics}
\eqn
%where $x(\cdot) \subset \mathcal{X} := \mathbb{R}^{n_x}$ is the state trajectory evolving from initial state $x$, $u(\cdot) \subset \mathcal{U} \subset \mathbb{R}^{n_u}$ is the control trajectory and $d(\cdot) \subset \mathcal{D} \subset \mathbb{R}^{n_d}$ is the disturbance trajectory, and $\mathcal{U}$ and $\mathcal{D}$ are convex sets constraining the respective inputs. 
where control and disturbance inputs $u$ and $d$ are drawn from convex sets $\mathcal{U} \subset \R^{n_u}$, $\mathcal{D} \subset \R^{n_d}$, 
and the control and disturbance functions $u(\cdot)$ and $d(\cdot)$ are assumed to be of the set of measurable functions $\mathbb U : [t,0] \mapsto \mathcal{U}$, $\mathbb D : [t,0] \mapsto \mathcal{D}$.
Assuming that the dynamics \eqref{Dynamics} is Lipschitz continuous in $(x,u)$ and continuous in $t$, there exists a unique trajectory $x(\cdot) \subset \mathcal{X} := \mathbb{R}^{n_x}$ of the system given initial state $x$, control function $u(\cdot)$, and disturbance function $d(\cdot)$. 

% \zgnote{Here, $\mathcal {X,U,D}$ are sets, denoting the control you take at a specific time right? and the corresponding trajectory should be a function: $u(\cdot): [t,T] \mapsto \mathcal U$ right? Then probably need another set denote mesurable function, i.e. $\mathbb U$ to denote these trajectory.}

\subsection{Hamilton-Jacobi Reachability Problem}\label{sec:HJR}
% Consider a system,
% \beqn
% \dot x(\tau) = f(x(\tau)), u(\tau), d(\tau), \tau) \quad \forall \tau \in [t, T],\: x(t)= x
% \label{Dynamics}
% \eqn

% \beqn
% ||f(x, u, d, t) - f(x', u, d, t)|| \leq L_f||x - x'|| \quad x, x' \in \mathbb{R}^{n_x}.
% % \label{Lipchitz}
% \eqn

To design a safe autonomous controller, HJ reachability solves for the optimal control that counters an adversarial disturbance in a differential game. Here, the control player's objective is to minimize the game cost while the disturbance player seeks to maximize it \cite{chow2017algorithm}. 
The game is defined by the cost functional
\beqn
P(x, u(\cdot), d(\cdot), t) = J(x(T)) + \int_t^T L(u(\tau), d(\tau)) d\tau ,
\label{GameCost}
\eqn
where $x(T)$ is the solution of (\ref{Dynamics}) at time $T$. The terminal cost $J:\mathbb{R}^{n_x} \rightarrow \mathbb{R}$ is a convex, proper, lower semicontinuous function chosen such that
\beqn
\begin{cases}
J(x) < 0 \:\: \text{ for } \:\:x \in \mathcal{T} \setminus \partial\mathcal{T} \\
J(x) = 0 \:\: \text{ for }\:\:x \in \partial\mathcal{T} \\
J(x) > 0 \:\: \text{ for } \:\:x \notin \mathcal{T}\\
\end{cases}
\label{InitialValue}
\eqn
where $\mathcal{T} \subset \mathcal X$ is a user-defined closed set representing the target to reach (or avoid, if the max/min are switched) and $\partial \mathcal{T}$ its boundary. The running cost $L:\mathbb{R}^{n_u} \times \mathbb{R}^{n_d} \rightarrow \mathbb{R}$ serves only to constrain the inputs and, thus, takes the form,
\beqn
L(u, d) = \mathcal{I}_\mathcal{U}(u) - \mathcal{I}_\mathcal{D}(d)
\label{RunningCost}
\eqn
where $\mathcal{I}_\mathcal{C}$ are the indicator functions of $\mathcal{U}$ and $\mathcal{D}$, defined such that
\beqn
\begin{cases}
\mathcal{I}_\mathcal{C}(c) = 0 & \text{ if } c \in \mathcal{C}\,;\\
\mathcal{I}_\mathcal{C}(c) = +\infty & \text{ otherwise.}
\end{cases}\label{Indicator}
\eqn
We have now defined the game such that for trajectory $x(\cdot)$ arising from a given $x$, $u(\cdot)\subset \mathcal{U}$, $d(\cdot)\subset \mathcal{D}$, $t$, and (\ref{Dynamics}), 
\beqn
P(x, u(\cdot), d(\cdot), t) \leq 0 \iff x(T) \in \mathcal{T}.
\label{GameMeaning}
\eqn

% where $p\in\mathbb{R}^{n_x}$ is the Lagrange multiplier of the dynamics constraint. In this paper we assume that $\forall t$ the Hamiltonian obtains a unique Nash equilibrium (saddle point) for the control and disturbance strategies (Isaac's condition). While restrictive, this assumption applies to a large class of systems, namely linearly-affine systems of the form,
% \beqn
% \dot x = f(x, t) + B_1(t)u + B_2(t)d.
% \label{DynamicsAffine}
% \eqn
% Isaac's condition guarantees that the game has value, i.e. that the lower and upper value of the game coincide and, thus, can be defined unanimously by
The function  $V:\mathbb{R} \times R \rightarrow \mathbb{R}$ corresponding to the optimal value of the game is defined as
\beqn
V(x,t) = \sup_{\mathit{d} (\cdot) \subset \Gamma(t)} \inf_{u(\cdot) \subset \mathcal{U}} P(x, u(\cdot), d(\cdot), t)
\label{GameValue}
\eqn
where $\Gamma(t)$ is the set of non-anticipative strategies defined in \cite{bacsar1998dynamic, darbon2016algorithms, chow2017algorithm, chow2019algorithm} and we assume Isaac's condition \cite{bacsar1998dynamic}.
This function is useful because by the same logic of (\ref{GameMeaning}),
\beqn
V(x,t) \leq 0 \iff x \in \mathcal{R}(\mathcal{T},t)
\label{ValueMeaning}
\eqn
where $\mathcal{R}(\mathcal{T},t)$, the BRS, is the set of all states which can be driven to the target for any bounded disturbance, 
\beqn
\begin{aligned}
\mathcal{R}(\mathcal{T},t) =\{x \:\:  | \:\:  \exists u(\cdot)\subset \mathcal{U} \:\: \forall d(\cdot)\subset \mathcal{D} \text{ s.t. } \\ x(\cdot) \text{ satisfies } (\ref{Dynamics}) \land x(T) \in \mathcal{T}\}.
\end{aligned}
\label{BRS}
\eqn
Notably, applying Bellman's principle of optimality to this time-varying value function $V$ leads to the following well known theorem.
\begin{mytheorem}
[Evans 84] \cite{evans1984differential}
Given the assumptions (2.1)-(2.5) in [Evan 84], the value function $V$ defined in (\ref{GameValue}) is the viscosity solution to the following Hamilton-Jacobi Partial Differential Equation,
\beqn
\begin{aligned}
\frac{\partial V}{\partial t}  + H(x, \nabla_x V, t) &= 0 \qquad \text{ on } \mathbb{R}^{n_x} \times [t,T], \\
V(x,T) &= J(x(T)) \:\: \text{ on } \mathbb{R}^{n_x} 
\end{aligned}
\label{HJPDE-V}
\eqn
where the Hamiltonian $H:\mathbb{R}^{n_x} \times \mathbb{R}^{n_x} \times [t,T] \rightarrow \mathbb{R}$ is defined as 
\beqn
H(x, p, t) = \min_{u \in \mathcal{U}} \max_{d \in \mathcal{D}} p \cdot f(x, u, d, t).
\label{Hamiltonian}
\eqn
\end{mytheorem}

To solve this PDE, therefore, yields the value function and corresponding BRS. Additionally, the value function can be used to derive the optimal control strategy for any point in space and time with:
\beqn
\begin{aligned}
u^*(x, t) = \arg \min_{u \in \mathcal{U}} \nabla_x V(x,t) \cdot g_1(x)u.
\end{aligned}
\label{HJoc}
\eqn

The main challenge of HJR lies in solving this PDE in (\ref{HJPDE-V}); DP methods propagate $V(x, t)$ by finite-differences over a grid of points that grows exponentially with respect to $n_x$ \cite{bansal2017hamilton}. In practice, this is computationally intractable for systems of $n_x \geq 6$ and constrained to offline planning.
    
\subsection{The Hopf Solution to HJ-PDE's}\label{subsec:hopf}

An alternative to the brute-force grid solving of $V(x,t)$ is the Hopf formula, which offers a solution to (\ref{HJPDE-V}) in the form of a space-parallelizeable optimization problem. 
First, we define $\phi(x,t):=V(x, T-t)$ to change the aforementioned final-value problem into an initial-value problem for simplicity. $\phi$ is now the solution of
\beqn
\begin{aligned}
- \frac{\partial \phi}{\partial t}  + H(x, \nabla_x \phi, t) &= 0 \qquad \text{ on } \mathbb{R}^{n_x} \times [0,t], \\
\phi(x,0) &= J(x) \:\: \text{ on } \mathbb{R}^{n_x} 
\end{aligned}
\label{HJPDE-phi}
\eqn
with Hamiltonian,
\beqn
H(x, p, t) = \max_{u \in \mathcal{U}} \min_{d \in \mathcal{D}} -p \cdot f(x, u, d, T-t).
\label{HamiltonianPhi}
\eqn
% and $f$ defined in (\ref{Dynamics}).

Note, for systems $f(x,u,d,t)=f(u,d,t)$ without state dependence, the Hamiltonian $H(x,p,t) = H(p,t)$ also lacks state-dependence and in this setting the following Hopf formula is available with limitation. This formula was conjectured in \cite{hopf1965generalized}, proved to be the viscosity solution in \cite{bardi1984hopf} and \cite{lions1986hopf} for $H(p)$, and generalized to some $H(t,p)$ in \cite{rublev2000generalized, kurzhanski2014dynamics}. Recently, \cite{darbon2016algorithms} devised a fast method of solving this formula and \cite{chow2019algorithm} conjectured a general form for $H(x,t,p)$.

\begin{mytheorem} [Rublev 2000] \label{Hopfthm}
We assume $J(x)$ convex and Lipchitz, and that $H(t,p)$ is pseudoconvex in $p$ and satisfies (B.i-B.iii) in [Rublev 00], then the minimax-viscosity solution \cite{subbotin1996minimax} of (\ref{HJPDE-phi}) is given by the time-dependent Hopf formula,
\beqn
\phi(x,t) = -\min_{p \in \mathbb{R}^{n_x}} \bigg\{ J^\star(p) - x \cdot p + \int_0^t H(p, \tau) d\tau \bigg\}
\label{HopfFormula}
\eqn
where $J^\star(p):\mathbb{R}^{n_x} \rightarrow \mathbb{R} \cup \{+\infty\}$ is the Fenchel-Legendre transform (i.e. convex-conjugate) of a convex, proper, lower semicontinuous function $J:\mathbb{R}^{n_x} \rightarrow \mathbb{R}$ defined by
\beqn
J^\star(p) = \sup_{x\in \mathbb{R}^{n_x}} \{ p \cdot x - J(x) \}.
\label{FL}
\eqn

Under these assumptions, the minimax-viscosity and viscosity solutions are equivalent when $H(s,p)$ is either convex or concave in $p$ for $s\in [0,t]$ \cite{rublev2000generalized, subbotin1996minimax}.
\end{mytheorem}
One may refer to \cite{lions1986hopf, rublev2000generalized, subbotin1996minimax, chow2019algorithm} for analysis and comparison of minimax-viscosity and viscosity solutions. Note, only the latter corresponds to the solution of the differential game when the two solutions diverge. 
If some space-time coupling is permitted, the two solutions may be forced to agree with temporal re-initialization of the value \cite{wei2013viscosity}, however, this requires caution and we do not attempt it in the current work. 
For general non-convex $H$, the question of when these solutions coincide remains open, however, like \cite{chow2019algorithm}, we observe some agreement in our results. 

One may note that there is an alternate form of the Hopf formula, called the Lax formula \cite{hopf1965generalized, lions1986hopf} which swaps the convexity assumption between $J$ and $H$. This permits a non-convex target but assumes a convex game. Moreover, depending on the method, assuming convexity of both $J$ and $H$ may be required to ensure rapid, global optimization. We discuss these nuances at the end of the section.

The strength of Hopf is that now we may compute (\ref{HJPDE-V} \& \ref{HJPDE-phi}) and solve our problem by solving a space-parallizeable optimization problem (i.e. $\phi(x,t)$ does not depend on $\phi(x',t'<t)$, unlike DP) \cite{darbon2016algorithms, chow2017algorithm, chow2019algorithm}, avoiding the so called \textit{curse of dimensionality}. However, to require a state-independent Hamiltonian limits this method to a specific class of systems. 
% Furthermore, we can not guarantee we have the solution which solves the game if the Hamiltonian is non-convex, which can occur based on the relative sizes of $\mathcal{U}$ and $\mathcal{D}$.

It was noted in \cite{kurzhanski2014dynamics, darbon2016algorithms, chow2017algorithm} that any linear time-varying system of the form
\beqn
\begin{aligned}
\dot x = A(t)x + B_1(t) u + B_2(t) d
\end{aligned}
\label{LinearStateIndependent}
\eqn
can be mapped to a state-independent system 
\beqn
\begin{aligned}
\dot z = \Phi(t)(B_1(t) u + B_2(t) d)
\end{aligned}
\label{LinearStateIndependent2}
\eqn
with the linear time-varying mapping $z(t):=\Phi(t; A)x(t)$ defined by the fundamental matrix $\dot \Phi = A(t) \Phi(t), \Phi(0) = I $. If $A(t) \equiv A$, then $\Phi(t; A) = \exp(-tA)$. After the change of variable, the Hamiltonian for $\phi$ becomes,
\beqn
\begin{aligned}
H_\mathcal{Z} (p, t) =& \max_{u \in \mathcal{U}} -p \cdot \Phi(T-t) B_1(T-t) u \\ 
& - \max_{d \in \mathcal{D}} -p \cdot \Phi(T-t) B_2(T-t) d .
\end{aligned}
\label{HamiltonianLinear}
\eqn
Since the mapping $\Phi$ is injective, $\phi_\mathcal{Z}(z,t) = \phi(x,t)$ and
\beqn
\phi_\mathcal{Z}(z,t) = -\min_{p \in \mathbb{R}^{n_x}} \bigg\{ J_\mathcal{Z}^\star(p) - z \cdot p + \int_0^t H_\mathcal{Z}(p, \tau) d\tau \bigg\}
\label{HopfFormulaZ}
\eqn
where $\phi_\mathcal{Z}(z, 0) = J_\mathcal{Z}(z(0)) = J(\Phi(T)x(T))$ and we define $T=0$, thus, $J_\mathcal{Z}(z) = J(x)$ and $J_\mathcal{Z}^\star(p) = J^\star(p)$ \cite{chow2017algorithm, donggun19iterativehopf}.
% \beqn
% \begin{aligned}
% z = z(0) = x(0) = x &\implies J_\mathcal{Z}(z) = J(x)  \\
% &\implies J_\mathcal{Z}^\star(p) = J^\star(p).
% % \label{eqn}
% \end{aligned}
% \eqn

Given the convexity of $\mathcal{U}$ and $\mathcal{D}$, the Hamiltonian in (\ref{HamiltonianLinear}) can be rewritten as the difference of two positively homogeneous Hamiltonians given by the convex-conjugates of the indicator functions of $\mathcal{U}$ and $\mathcal{D}$ \cite{chow2017algorithm},
\beqn
\begin{aligned}
H_\mathcal{Z} (p,t) = \mathcal{I}_\mathcal{U}^\star(R_1(t) p) - \mathcal{I}_\mathcal{D}^\star(R_2(t) p), \\ R_i(t):=-B_i(T-t)^\dagger \Phi(T-t)^\dagger .
\end{aligned}
\label{HamiltonianIndicator} 
\eqn
This allows rapid and efficient computation of (\ref{HopfFormulaZ}) that has been observed to scale linearly with $n_x$ for a fixed $t$ \cite{chow2017algorithm} and demonstrated in online guidance with naive linearizations \cite{Kirchner_2018}. Notably, when $\mathcal{U}$ and $\mathcal{D}$, are constrained by norms, these functions have solutions, namely the dual norms. For example, given $Q \succeq 0$, consider two popular constraints \cite{chow2017algorithm}, the ellipse and box, and the corresponding $\mathcal{I}^\star$:
\beqn
\begin{aligned}
 \mathcal{C}_{\mathcal{E}}(Q) &:= \{c\:\:|\:\:||c||_{Q^{-1}} \leq 1\} \implies \:\: \mathcal{I}_\mathcal{C}^\star(\cdot) = ||\cdot||_{Q}, \\
 \mathcal{C}_{\mathcal{R}}(Q) &:= \{c\:\:|\:\:||Q^{-1}c||_\infty \leq 1\} \implies \:\: \mathcal{I}_\mathcal{C}^\star(\cdot) = ||Q\cdot||_1.
\end{aligned}
\label{InputConstraint} 
\eqn
% where $||\cdot||_Q = ||\Sigma V^T \cdot||_2$ is the elliptical norm described by $Q^{-1}$ and $||Q\cdot||_\infty$ the rectangular norm. 
For these set types, we note the following conditions to certify the convexity of the Hamiltonian (\ref{HamiltonianIndicator}) for a state-independent system in (\ref{LinearStateIndependent2}). If $\mathcal{U}$ and $\mathcal{D}$ are defined by the same set type in (\ref{InputConstraint}) with $Q_u$ and $Q_d$ respectively, and $B_1 = B_2$, then convexity of the Hamiltonian is given when $Q_u \succeq Q_d$ i.e. when the control authority ``exceeds" the disturbance authority. When $B_1 \neq B_2$, convexity of the Hamiltonian is given when $B_1 Q_u B_1^\dagger \succeq B_2 Q_d B_2^\dagger$ or $[Q_u B_1^\dagger \mathbf{1}]_j > [Q_d B_2^\dagger \mathbf{1}]_j \forall j$ when the input sets $\mathcal{U}$ or $\mathcal{D}$ are both defined by $\mathcal{C}_{\mathcal{E}}$ or $\mathcal{C}_{\mathcal{R}}$ respectively.

We may note, from the definition of the Fenchel-Legendre (FL) transform, (\ref{HopfFormula}) may be rewritten as \cite{lions1986hopf}
\beqn
\phi(x,t) = \bigg(J^\star + \small\int_0^t H\bigg)^\star (x).
\label{FLphi}
\eqn
This illustrates that the value function is itself an FL transform and from the well known property of the FL transform \cite{lions1986hopf}, $\nabla_x \phi(x,t)$ is the minimum argument of (\ref{HopfFormula}). Apart from being a perspective for understanding the Hopf solution, we may use this fact to compute the optimal control strategy in (\ref{HJoc}) immediately from the numerical solution method. 

% \begin{mytheorem}[Warm-Starting Hopf]
%     We assume $J(x)$ is Lipchitz with constant $L$ and the conditions of Thm \ref{Hopfthm}, then for $x, x' \in \mathbb{R}^{n_x}$ and $p^*, p^{*'}$ defined as the corresponding $\arg \min$ of (\ref{HopfFormula}) respectively, 
%     \beqn
%     ||p^* - p^{*'}|| < 2L.
%     \eqn
%     \textit{Proof}:
%     Let $h = ||x - x'||$. By the fact given in (\ref{FLphi}) and the well-known property of the FL transform \cite{lions1986hopf},
%     \beqn 
%      ||p^*|| = || \nabla \phi(x,t) ||
%      = \lim_{h \rightarrow 0} ||\phi(x,t) - \phi(x',t)||/h \\ < \lim_{h \rightarrow 0} Lh/h  = L.
%     \eqn
%     By the same logic $||p^{*'}|| < L \implies ||p^* - p^{*'}|| < 2L$ \QED
% \end{mytheorem} 
\noindent 

Moreover, we may use this fact for initializing space-neighboring Hopf optimization problems. Considering the well-known Lipchitz continuous quality of $\phi$, for $x, x' \in \mathbb{R}^{n_x}$ and $p^*$ and $p^{*'}$ defined as the corresponding $\arg \min$ of (\ref{HopfFormula}),
\beqn
||p^* - p^{*'}|| < ||p^* - 0|| + ||p^{*'} + 0|| < 2L ||x - x'||
\eqn
Thus, if we are solving independent Hopf problems over a space sequentially, using the minimizing argument of the last problem initializes the optimization within a $2L$ neighborhood of the optimum. In practice, we observe this accelerates convergence by up to 10-fold or greater in high dimensional problems. 

Although more scalable than DP (due to the \textit{curse of dimensionality}), the numerical optimization of the Hopf formula is a non-trivial problem discussed at length in \cite{darbon2016algorithms, chow2017algorithm, Kirchner_2018, chow2019algorithm} (often referred to as the \textit{curse of complexity}). For the examples in this work, we found that the Coordinate-Descent (CD) \cite{bezdek1987local, boyd2004convex} and the Alternating Direction Method of Multipliers (ADMM) \cite{gabay1976dual, boyd2011distributed} algorithms tailored to the Hopf formula in \cite{chow2017algorithm} were suitable for our problems. These are implemented in our codebase \href{https://github.com/UCSD-SASLab/HopfReachability}{\textit{HopfReachability.jl}} and available for public use.

Notably, however, these algorithms impose additional assumptions for solving the Hopf formula: first, for quickly computing the FL transformation (in both CD or ADMM), and second, for computing the proximal mappings (in ADMM). The first condition refers to the need to evaluate $J^*$ in \ref{HopfFormulaZ} at each iteration, which may require a sub-step optimization if it does not have a closed form (e.g. irregular/piece-wise convex). Recall, that the Hopf formula assumes convexity of $J$, thus the FL transformation must have a unique optimum but finding this optimum might be slow for high dimensional problems, delaying each iteration of the numerical optimization. Thus, in this work, we limit our problems to targets described by norms, which have closed-form FL transformations (given in \ref{InputConstraint}) \cite{boyd2004convex}. Secondly, if one desires to use ADMM, known to be efficient and dimension-robust \cite{boyd2011distributed}, one needs to compute the proximal mappings associated with $\mathcal{U}$ and $\mathcal{D}$ at each iteration \cite{chow2017algorithm, chow2019algorithm} that also amounts to sub-step optimization. Like before, although we have assumed convexity of each set, irregular convex geometries in high dimensions can slow the convergence to the unique optimum, delaying each iteration of the numerical optimization. Thus, we limit the current work to control and disturbance sets constrained by norms  (\ref{InputConstraint}).% If more sophisticated methods were employed in the sub-steps, one might accelerate these algorithms and allow for a broader class of reachability problems (and Koopman methods when joined, see below), however, we leave this for future work.

In summary, the use of the Hopf formula requires several theoretical and practical assumptions that are important to consider. First, we assume $J$ and $H$ convex for the formula to give the value function associated with the game. This requires that our reachability target $\mathcal{T}$ must be convex and the control set $\mathcal{U}$ must \textit{exceed} the disturbance set $\mathcal{D}$ (\ref{InputConstraint}). Second, we assume these convex sets $\mathcal{T}$, $\mathcal{U}$, $\mathcal{D}$ are given by norms so that the convergence of the numerical optimization is rapid. These requirements have vital implications for the available lifting functions used in the Koopman methods and the definition of a target in the lifted space.

\subsection{Koopman Theory}\label{sec:Koop}

Solving a Hopf reachability problem requires access to a linear model representing the system dynamics globally across the state space. Among the options for linearizing nonlinear dynamic systems, 
%We would like to apply dimension-robust Hopf theory to high-dimensional, nonlinear systems with a highly accurate linearization to approximate BRSs and synthesize safe control. Moreover, we would like a non-local linearization method due to the BRSs potentially global expansion.
Koopman theory is known for outperforming other methods in producing accurate, non-local linearizations \cite{proctor2014dynamic, korda2018linear, yeung2019learning}.

Consider the discretized mapping of a nonlinear system,
$
x(t_{i+1}) = F(x(t_i))$.
The Koopman operator \cite{koopman1931hamiltonian, mezic2021koopman} $\mathcal{K}:\mathcal{F} \rightarrow \mathcal{F}$ is defined as
% \beqn
$\mathcal{K} g:=g \circ F$,
% \label{KoopDef}
% \eqn
where $\mathcal{F}$ is the collection of all functions that form an infinite Hilbert space, often called observables, and $g \in \mathcal{F}:\mathcal{X} \rightarrow \mathbb{R}$ \cite{mezic2021koopman, CompKoop}. By definition, the operator has the property,
\beqn
(\mathcal{K}g)(x(t_i)) = g(F(x(t_i))) = g(x(t_{i+1}))
\label{mezic2021koopman, KoopProperty}
\eqn
and we assume this holds for a finite space such that $Kg \in \mathcal{G}$ \cite{mezic2021koopman, CompKoop}, where $\mathcal{G}$ is an invariant subspace of the Koopman space. We assume this space is spanned by a finite basis of observables $\{\psi_1(x),\: \dots, \psi_{n_k}(x) \}$, thus, $K \in \mathbb{R}^{n_k \times n_k}$ is a finite matrix. We call the vector-valued mapping to the concatenated basis $\Psi(x):= [\psi_1(x),\: \dots, \psi_{n_k}(x)]^\dagger$ the lifting function. Then, in practice, for a data set of points $X(t_i)$ and their one-step evolution $X(t_{i+1})$, $K$ can be approximated from the least-squares problem,
\beqn
\min_K || \Psi(X(t_{i+1})) - K\Psi(X(t_{i})) ||^2.
\label{KoopmanFitting}
\eqn 
%Despite that Koopman and Von Neumann did not consider a system with external inputs, several group 
Recent works have found this theory can be extended to systems with external inputs \cite{proctor2014dynamic, korda2018linear, mezic2021koopman, kaiser2021data,CompKoop} such that for $x(t_{i+1}) = F(x(t_i),u(t_i),d(t_i))$. We use the Koopman control form \cite{ proctor2014dynamic, korda2018linear},
\beqn
g(x(t_{i+1})) \approx Kg(x(t_{i})) + L_1 u(t_i) + L_2 d(t_i),
\label{KoopControl}
\eqn
where $L_1$ and $L_2$ are the Koopman control and disturbance matrices. We note that several groups have observed improved accuracy in prediction and control when lifting the inputs (in addition to the state), e.g. with $u_g := \Psi(u)$ or $u_g := 
 \Psi(x, u)$, however, the use of the Hopf formula requires several assumptions that constrain this direction (Sec.~\ref{subsec:hopf}), thus, we leave this formulation for future work. The system  \eqref{KoopControl} is similarly fit with control and disturbance data $U(t_i)$ and $D(t_i)$ from a least-squares problem,
% \beqn
% \min_{K, L_1, L_2} || \Psi(X(t_{i+1})) - (K\Psi(X(t_{i})) + L_1 U(t_i) + L_2 D(t_i)) ||.
% \label{KoopmanControlFitting}
% \eqn 
\beqn
\min \left\lVert \Psi(X(t_{i+1})) - \begin{bmatrix} K & L_1 & L_2 \end{bmatrix} \begin{bmatrix} \Psi(X(t_i)) \\ U(t_i) \\ D(t_i) \end{bmatrix} \right\rVert^2.
\label{KoopmanControlFitting}
\eqn 
We note that there are varying technical approaches to solving (\ref{KoopmanFitting}) and (\ref{KoopmanControlFitting}) and we refer readers to \cite{korda2018linear, mezic2021koopman, CompKoop, bruder2019modeling} in general, and to \cite{dynamicslab_pykoopman, EthanJamesLew_AutoKoopman} for the specific details of the methods used in this work. 

Finally, we define a ``lowering'' function 
$\widetilde{\psi}:\mathcal{G} \rightarrow \mathcal{X}$ for mapping the Koopman predicted evolution in the true space, as in \cite{bruder2019modeling, proctor2014dynamic, korda2018linear, yeung2019learning}, with the only requirement that $\widetilde{\psi}(\Psi(x)) \approx x$; this can be a simple projection to the true space when the identity function is included in the lifting dictionary $\Psi$, as in our examples. Note, the inclusion of the identity function in the lifting function yields theoretical inconsistencies when the system has multiple fixed points, with respect to topological conjugacy \cite{lan2013linearization, brunton2016koopman}, or when the system includes linearly-injected control, with respect to the chain-rule \cite{lan2013linearization, bakker2019koopman}. Nonetheless, often accuracy is sufficient within a large region of interest and by nature, allow simple interpretation of the Koopman space, and we include this possibility in our work.

Notably, this lowering function is not injective; due to the approximate nature of (\ref{KoopControl}) (e.g. arising from truncation error, fitting error or control error), 
\begin{align}
    K\Psi(x) + L_1u + L_2d &\in \mathcal{G}=\text{Span}(\Psi), \quad  \text{but} \\ K\Psi(x) + L_1u + L_2d &\notin \text{Image}(\Psi),
\label{KoopmanApprox}
\end{align}
where $\text{Span}(\Psi):=\{g| g = c\cdot\Psi(x), c\in \mathbb{R}^{n_k}\}$ ($= \mathcal{G}$), and $\text{Image}(\Psi):=\{g|g=\Psi(x)\}$, the manifold of lifted true states. Hence, trajectories in the approximate Koopman space that start at points lifted from the true space may be driven to points which do not have a correspondence to true points, as discussed in \cite{bruder2019modeling}. Yet, we might like to interpret these trajectories and, thus, we allow the lowering function to condense infinitely many Koopman trajectories into the same true trajectory, like in \cite{proctor2014dynamic, brunton2016koopman, korda2018linear, yeung2019learning}. Although possible, the lowering function is not used in the fitting, however, it guides the definition of a target in the Koopman space in Sec.~\ref{sec:koopman_hopf} that accounts for the trajectories in discussion, which proved the most accurate for approximating the true reachable set.

\section{Koopman-Hopf Reachability}\label{sec:koopman_hopf}

% \subsection{Reachability in Lifted Spaces}\label{sec:lifted_reach}

We now seek to approximate the value function and BRS (from Sec.~\ref{sec:HJR}) of our differential game by solving the Hopf formula (\ref{HopfFormulaZ}) with approximate linear dynamics derived from Koopman methods. We will first show how to define the target set in the Koopman space, which we call the ``lifted target" for its relation to the lifting function. We will then discuss the Hopf requirements that this lifted target must satisfy and the lifting functions that are well-suited to generate a satisfactory lifted target. Finally, we describe how to solve the resulting Hopf reachability analysis using the lifted target set and dynamics.

\subsection{Defining the Lifted Target Set}

\begin{figure}
    \centering
    \includegraphics[width=\linewidth]{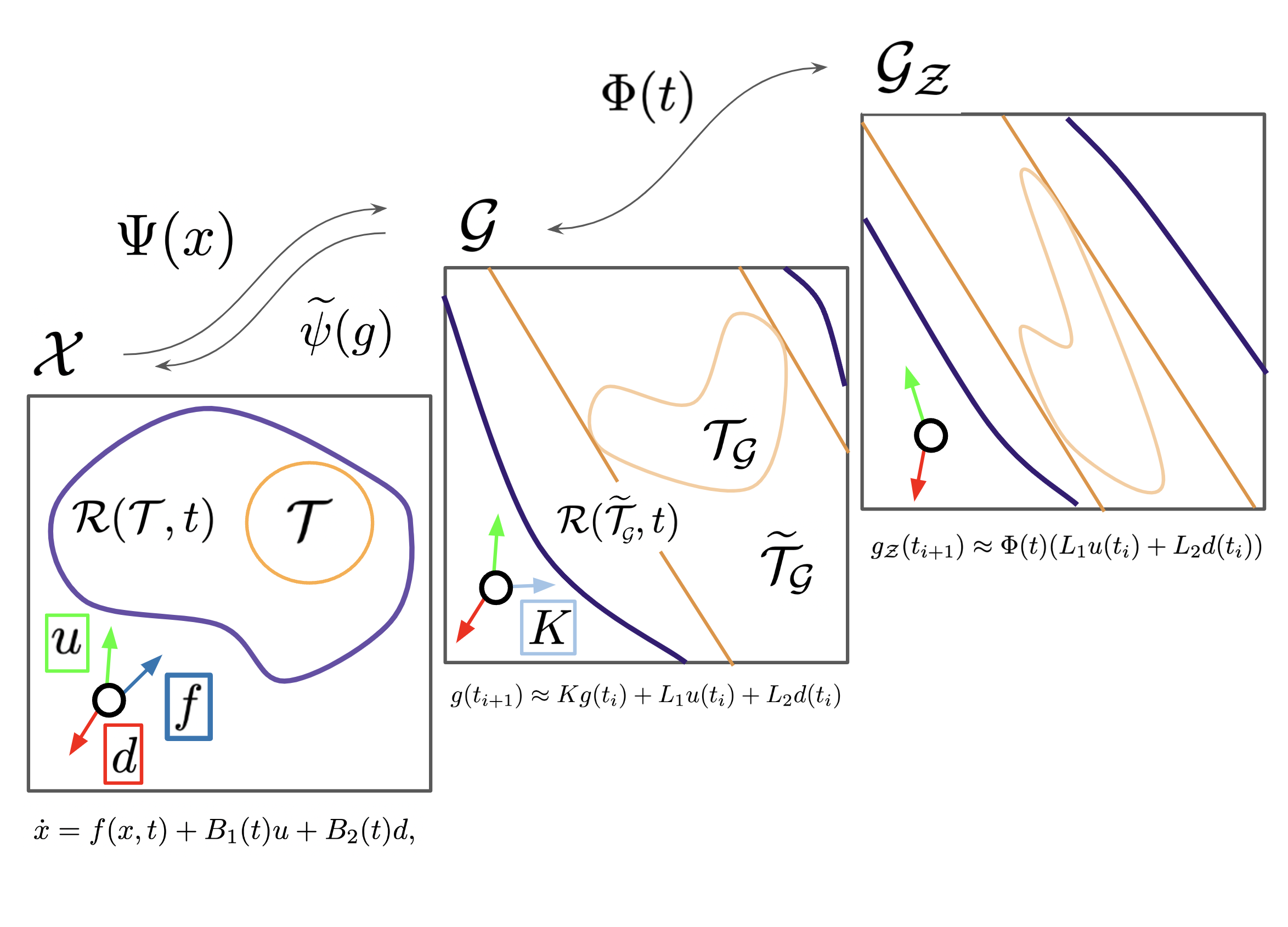}    
    \caption{\textbf{Sketch of the Involved Spaces \& Maps} This illustration depicts the three spaces -- the space of the true system $\mathcal{X} \subset \mathbb{R}^{n_x}$, the Koopman space $\mathcal{G} \subset \mathbb{R}^{n_k}$, and the state-independent Koopman space $\mathcal{G}_\mathcal{Z} \subset \mathbb{R}^{n_k}$ -- traversed in this work and the mappings between them: the lifting function $\Psi: \mathcal{X} \rightarrow \mathcal{G}$, the (invertible) fundamental matrix $\Phi: ([0, t], \mathcal{G}) \leftrightarrow \mathcal{G}_\mathcal{Z}$, and the lowering function $\tilde \psi: \mathcal{G} \rightarrow \mathcal{X}$. This is done to arrive at a space where we can solve the Hopf formula, but requires defining a target ($\mathcal{T}_\mathcal{G}$ or $\tilde{\mathcal{T}}_\mathcal{G}$) in the ``lifted" space that accurately represents the original reachability problem and satisfies the Hopf formula requirements.}
    \label{fig:Mapping_Graphic}\vspace{-1em}
\end{figure}

A target $\mathcal{T}$ defined in Sec.~\ref{sec:HJR} can be lifted directly to the Koopman space such that the lifted target $\mathcal{T}_\mathcal{G}$ is defined as the image of the target $\mathcal{T}$ under $\Psi$:
\beqn
\mathcal{T}_\mathcal{G} := \{ g \:\: |  \:\: g = \Psi(x), \: x \in \mathcal{T}\}.
\label{KoopTarget}
\eqn
However, given that we may not have an exact linearization as described in Sec.~\ref{sec:Koop}, we also propose an augmented lifted target $\widetilde{\mathcal{T}}_\mathcal{G}$ as the preimage of the target $\mathcal{T}$ under the lowering function $\widetilde{\psi}$:
\beqn
\widetilde{\mathcal{T}}_\mathcal{G} := \{ g \:\: | \:\: \widetilde{\psi} (g) \in \mathcal{T}\}.
\label{KoopTargetApprox}
\eqn
This captures the Koopman trajectories which might be driven out of the set of the lifted true states, i.e. the image of the states under the lifting function, to states in the Koopman space for which $\nexists x \in \mathcal{X}$ such that $g = \Psi(x)$ (see Sec.~\ref{sec:Koop}). These two sets are visualized in Figure~\ref{fig:Mapping_Graphic}.

If the identity function is included in the lifting function s.t. $\Psi(x) := [x, \psi_1(x), ...]$ and $\tilde \psi:=\text{Proj}_\mathcal{X}$, the augmented lifted target is given by the linear combination:
\begin{equation}
    \widetilde{\mathcal{T}}_\mathcal{G} =  \{ g \in \mathcal{G}\:|\: g = [1, c] \cdot \Psi(x),\: x \in \mathcal{T},\: c \in \mathbb{R}^{n_k-n_x}\}.
\end{equation}
This corresponds to the infinite extrusion of the true target and, thus, if the true target is given by (\ref{InputConstraint}) with $A \succeq 0$, then
\beqn
\widetilde{\mathcal{T}}_{\mathcal{G}, \mathcal{C}} = \mathcal{C}(\hat A), \:\: \hat A:=\begin{bmatrix}A & 0 \\ 0 & 0\end{bmatrix} \in \mathbb{R}^{M \times M}
\label{PolynomialTargetApprox}
\eqn
is an infinite-length cylinder or rectangular prism. In accordance with the true system, it might be prudent to under-approximate the boundless extrusion to allow some evolution from the lifted true states but limit grandly ``abstract" flows, potentially the result of poor Koopman approximation. Moreover, in practice we also find that relaxations of the infinite set, given by defined by $\mathcal{C}(Diag[A, \eps I])$ for $\eps >> 1$, accelerate convergence of the Hopf objective (Sec.~\ref{sec:results}).

%On the contrary, we are only concerned with finding the value of the states $\phi_\mathcal{G}(g,t)$ in the Koopman space such that $\exists x$ mapping to $g = \Psi(x)$. The value of states off of the true-state manifold might be the result of backwards-propagating a trajectory which may not ever have a real initial condition.

% \subsection{Lifting the target set and dynamics}
\subsection{Conditions for the Lifted Target} \label{subsec:Lifted_target_conditions}

The choice of lifting function impacts the shape of
the sets $\mathcal{T}_\mathcal{G}$ and $\widetilde{\mathcal{T}}_\mathcal{G}$. For the Hopf formula to be a solution to the game, the cost function $J$ and target set $\mathcal{T}$ must be convex\footnote{One could use the Lax formula to relax this assumption, however, it would ultimately require solving non-convex proximal maps in the numerical optimization.}.%, which we leave to future work.}. 

Convexity of the lifted target can be enforced by either,
\begin{enumerate}
    \item using a lifting function $\Psi(x)$ that preserves the convexity of a set in the region of the target, or 
    \item using an inner (for reach) or outer (for avoid) convex enclosure, e.g. ball or convex-hull of $\Psi(\mathcal{T})$.
\end{enumerate}
% An important requirement for utilization of the Hopf solution is convexity of $J(x)$. This is true if the Target $\mathcal{T}$ is convex, thus, we need $\mathcal{T}_\mathcal{G}$ convex. We propose a couple solutions: \zgnote{Here you are saying that the target is convex is the sufficient condition for a convex cost right? Is this always true? Or maybe you want to say if the target is convex, we can always define a convex cost?}
% \begin{itemize}
%     \item Take the inner (for reach) or outer (for avoid) ball or convex-hull of $\Psi(\mathcal{T})$
%     \item Use the Lax formula [YTC19], which swaps the convexity assumption between $J$ and $H$
%     \item Use lifting function $\Psi$ that are convex
% \end{itemize}

Both methods have nuances worth noting. Using a lifting function that preserves convexity is challenging, given that neither the difference nor concatenation of convex functions generally preserves the convexity of mapped sets. Thus, the notion of a convex basis is ill-defined, and to the best of the authors' knowledge, there exists no necessary and few sufficient conditions on functions that preserve the convexity of sets. This poses a significant challenge in the certification and construction of specialized lifting functions. Furthermore, we found that lifting functions that preserve convexity generally had greater residual error in the fitting of the Koopman dynamics. 
% Note, to use a lifting function of mixed quality and limit analysis where convexity is preserved limits target redefinition with a fixed Koopman lift and may perhaps subject the fit dynamics near the target to higher error for the aforementioned reasons. 

On the other hand, convex-hull methods suffer from the \textit{curse of dimensionality}, but the hull computation is only be performed once per problem formulation (for a static reach or avoid target). Note that this method of ``convexifying'' a lifted target from a non-convex lifting function can be overly conservative for highly non-convex lifting functions. However, this approach is typically preferred, as convex-enclosure methods can be adapted to the broad library of existing lifting functions tried in the literature.

For either case, although the Hopf formula assumes only the convexity of $J$ (and thus $\mathcal{T}$), to rapidly optimize the objective imposes further conditions regarding convex regularity (Sec.~\ref{subsec:hopf}) on a potential lifted target. %With respect to lifting functions, this requires that they not only preserve convexity (in the target region) but also regularity, which is fairly restrictive. As mentioned before, 
This regularity is not strictly required to solve the Hopf formula, however, it accelerates the sub-step optimization for online usage. %The proposed framework is amenable to varying the optimizer for improving the sub-step optimization and relaxing this condition, but we leave this to future work. 

This paper focuses on convex-enclosures of the lifted target with a norm-constrained set (\ref{InputConstraint}) or $\widetilde{\mathcal{T}}_\mathcal{G}$ which is convex for the true targets given by (\ref{InputConstraint}). Ultimately, the method of convex enclosures was chosen because it is simple to implement and flexible to existing Koopman linearizations. The problems of computing the maximum or minimum bounding ellipse or box are well-known in convex geometry with several available algorithms \cite{john2014extremum, ben2001lectures}. 
Note, although we may use a convex enclosure of the lifted target to over/under approximate the reachable set in the Koopman space (for avoid/reach problems respectively), these are not guaranteed to yield a strict subset/superset of the true reachable set if the Koopman lift of the true dynamics is not exact. %due to the approximate nature of the Koopman dynamics. 
%, and doesn't sacrifice the accuracy of the lifted dynamics - and offers good over/under approximations of the true reachable set.

\subsection{Potential Lifting Functions}

We demonstrate the proposed method with two simple lifting functions:
\begin{enumerate}
    \item the identity mapping, $I:\mathbb{R}^{n_x} \rightarrow \mathbb{R}^{n_x}$ (i.e. \cite{proctor2014dynamic})
    \item the multivariate polynomial of degree $l$, 
$P_l:\mathbb{R}^{n_x} \rightarrow \mathbb{R}^{M}, M = {n_x + l \choose l}  - 1$.
\end{enumerate}
Both cases have been extensively studied in Koopman literature; the former corresponds to dynamic-mode-decomposition with control (DMDc) \cite{proctor2014dynamic} while the latter is a case of extended-DMD with control \cite{korda2018linear, kaiser2021data}. 

The identity mapping is trivial and thus
\beqn
I(\mathcal{X}) = \mathcal{X} \implies \mathcal{T} = \mathcal{T}_\mathcal{G} = \widetilde{\mathcal{T}}_\mathcal{G}
\label{IdentityTarget}
\eqn
(assuming $\widetilde{\psi} = I$). Clearly, our lifted target will be convex if our initial target is convex and computing the FL transform of it will be identical to that of the original target. While this yields a trivial lifted target, it is a fairly limited Koopman method with higher residual error. 

%Although, we find it successful for one example of Koopman-based high-dimensional nonlinear control (\ref{sec:ctrl}), the 
The polynomial mapping $P_l(x)$ is more expressive and has been shown to produce more accurate linearizations of dynamics. However, it maps convex sets to non-convex sets, requiring an enclosure to yield a convex lifted target. 
% The benefit of using the polynomial mapping over a more complicated lifting function is that the inner or outer ellipse of the lifted target may be found analytically, skirting the need for volume-bounding algorithms which suffer in high-dimensionality.

Notably, both lifting functions include include the identity map. The presence of the identity function in the lifting function makes the definition of the lowering function and the augmented lifted target trivial. However, as discussed earlier, this Koopman system will be inconsistent when used to approximate systems with multiple fixed points and with control \cite{lan2013linearization, brunton2016koopman, bakker2019koopman}, ultimately limiting the global quality of the linearization. 

\subsection{Conditions for the Lifted Dynamics}

In addition to the target convexity, to guarantee that the Hopf formula solves the viscosity solution coinciding with the game - and numerical convergence to the global optimum - convexity of the Hamiltonian is sufficient. If the control and disturbance enter the Koopman space as in (\ref{KoopControl}), the sets will remain convex, however, the (fit) $L$ matrices may alter the relative sizes of the sets such that the Hamiltonian of the Koopman system is not convex. 

In the case of reachability problems for systems (\ref{Dynamics}) of the specific form $h_1(x) = h_2(x)$, e.g. when disturbance perturbs control actuation only, we may assume a controlled Koopman system (\ref{KoopControl}) with a corresponding form with $L_1=L_2$. Then, convexity of the Hamiltonian is given by $Q_u \succeq Q_d$ (see after (\ref{InputConstraint})) which will not be altered in the Koopman fitting process. However, for general Koopman systems where $L_1 \neq L_2$ and they are fit independently, the conditions $L_1 Q_u L_1^\dagger \succeq L_2 Q_d L_2^\dagger$ or $[Q_u L_1^\dagger \mathbf{1}]_j > [Q_d L_2^\dagger \mathbf{1}]_j \forall j$ must be checked (depending on the set type) to certify convexity of the Hamiltonian. In practice, however, the non-convex Hopf formula (minimax-viscosity solution) can be solved with a powerful optimizer like ADMM and matches the viscosity solution approximately.% (Fig.~\ref{fig:BRS_pannel}). 

Finally, note that if the Koopman system were defined with lifted inputs $u_g := \Psi_u(x,u)$ and $d_g := \Psi_d(x,d)$, the functions $\Psi_u$ and $\Psi_d$ would need to be treated with the same caution as in Sec. \ref{subsec:Lifted_target_conditions} to certify the convexity of the Hamiltonian.%, however, we leave this for future work. 

\subsection{Koopman-Hopf Procedure Summary}

In total, the proposed method involves the following steps, listed in Algorithm~\ref{alg:KHP}. First one must choose a lifting function that yields a suitable Koopman system (low residual error in (\ref{KoopmanControlFitting})) and lifted target (satisfying Sec.~\ref{subsec:Lifted_target_conditions}). Next, points of interest -- such as a set for interpolating the BRS or a singleton corresponding to an online control program -- must be lifted to the Koopman space and then mapped to the state-independent version of it with the fundamental map (and fit $K$), corresponding to (\ref{LinearStateIndependent2}). Note, a graphical overview of these involved spaces and mappings can be seen in Fig.~\ref{fig:Mapping_Graphic}. Finally, the value at each of these lifted points is solved by optimizing the Hopf formula in (\ref{HopfFormulaZ}) and is attributed to the true point that was lifted, approximating the true value. For optimization in this work, ADMM is paired with CD (see \href{https://github.com/UCSD-SASLab/HopfReachability}{\textit{HopfReachability.jl}} for parameter details). If interested in the approximate optimal control and disturbance at this point, one may compute these with the minimizing argument (equivalent to the approximate gradient, see (\ref{FLphi})) inserted in the relation in (\ref{HJocHopf}). If interested in the approximate BRS, one may estimate the zero-level set from the approximate values (e.g. by interpolation).

\begin{figure}
\vspace{-0.15in}
\begin{algorithm}[H]
\caption{Koopman-Hopf Procedure}
\label{alg:KHP}
\begin{algorithmic}[1]
\REQUIRE For a given system $f(x,u,d)$, $\mathcal{U}$ \& $\mathcal{D}$ (\ref{Dynamics}), \\
Target: $\mathcal{T} \rightarrow J(x)$ (\ref{InitialValue}) \\
Time(s) to solve: $T-t$ \\
State(s) to solve: $X^s$ \emph{\color{gray} (e.g. grid for BRS, one for control)}
\item[]

\STATE Choose $\Psi(x)$ and $\mathcal{T}_\mathcal{G} \rightarrow J_\mathcal{G}(g)$ satisfying Sec.~\ref{subsec:Lifted_target_conditions}
\STATE Fit $(K, L_1, L_2)$ from system trajectories (\ref{KoopmanControlFitting})
\STATE Lift $X^s$ by $\mathcal{G}_\mathcal{Z}^s = \Phi(T-t; K) \Psi(X^s)$ (\ref{LinearStateIndependent2})
\item[]

\FOR{$g_\mathcal{Z} \in \mathcal{G}_\mathcal{Z}^s$ \emph{\color{gray} in parallel}}
    \STATE Solve $\phi_\mathcal{Z}(g_\mathcal{Z}, t)$ by optimization of (\ref{HopfFormulaZ}) with $J_\mathcal{G}^*$ (\ref{FL}) and $H_{\mathcal{G}_\mathcal{Z}}$ (\ref{HamiltonianLinear})
    \STATE $V_\mathcal{G}(g,T-t)$, $\nabla_{g_\mathcal{Z}} \phi_\mathcal{Z}(g_\mathcal{Z}, t) = \phi_\mathcal{Z}(g_\mathcal{Z}, t)$, $\hat p^*$ 
    \STATE $\hat V(x, T-t) := V_\mathcal{G}(g_\mathcal{Z},T-t)$ for $g_\mathcal{Z} = \Phi(T-t; K) \Psi(x)$
    % \STATE \emph{\color{gray} warm starting???}
    \IF{solving strategies}
        \STATE $\hat u^*(x,T-t), \hat d^*(x,T-t)$ are computed with $K$, $L_1$, $L_2$ and $\partial_p H_{\mathcal{G}_\mathcal{Z}}(\nabla_{g_\mathcal{Z}}\phi_\mathcal{Z}(g_\mathcal{Z}, t), t; \mathcal{U}, \mathcal{D})$  (\ref{HJocHopf})
    \ENDIF
\ENDFOR
\item[]

\IF{solving BRS}
    \STATE estimate $\hat V(x, T-t)=0$ for $x \in X$ from $\hat V(x, T-t)$ for $x \in X^s$
\ENDIF
\RETURN Approx. BRS $\hat V(x, T-t)=0$ \\ and/or Approx. strategies $\hat u^*(x,T-t), \hat d^*(x,T-t)$ \\
\end{algorithmic}
\end{algorithm}
\vspace{-10mm}
\end{figure}
% Requirements
% \begin{enumerate}
%     \item For hard guarantees, need exact linearization and maintain convexity. Only in specific cases, i.e. with convex lifting function and with disturbance on control only can we guarantee convexity.
%     \item If relax these requirements, lose guarantees but using a powerful solver which is known to perform well in non-convex cases, we gain approximations for high-dim nonlinear systems. Can be used as-is or to guide e.g. training.
%     \item how to lift target set / function to high-dim space and back.
% \end{enumerate}

% \begin{figure*}[t]
%     \centering
%     \includegraphics[width=\linewidth]{Glycolysis_experiment_2.png}    
%     \caption{\textbf{Comparison of Koopman-based Controllers in a Glycolysis System} Three controlled evolutions, using the same Koopman lift, of the 10D glycolysis model with the same disturbance trajectory and initial condition. `Auto' signifies the disturbed autonomous system. The bottom pannels plot the three controlled states in the model that can be controlled. Koopman-Hopf controller ("Hopf" above) amplifies the cycle of the phosphate, glucose and ATP states by oscillating the total concentrations of NAD+/NADH and ATP/ADP to achieve the target in a manner that translates to the nonlinear system.}
%     \label{fig:glycolysis}\vspace{-1em}
% \end{figure*}

\section{Results}\label{sec:results}

All results are computed using our codebase \href{https://github.com/UCSD-SASLab/HopfReachability}{\textit{HopfReachability.jl}}, an open-source package designed for solving 2-player linear differential games.

\subsection{Approximate BRS of the Slow Manifold}

We begin by considering the well-known slow manifold system (see \cite{brunton2016koopman, korda2018linear} for detailed Koopman analysis),
\begin{align}
    \dot{x} =  
    \begin{bmatrix} \mu x_1 \\ \lambda (x_2 - x_1^2) \end{bmatrix},
    \label{SlowManifold}
\end{align}
with $\mu, \lambda := -0.05, -1$. The autonomous system is famous for being exactly linearizeable with the lifting function $\Psi_{SM}(x):=[x_1, x_2, x_1^2]^\dagger$ such that $g \in \text{Image}(\Psi_{SM}) \subseteq \mathcal{G}$, 
\begin{align}
    \dot{g} = 
    \begin{bmatrix} \mu & 0 & 0 \\ 0 & \lambda & - \lambda \\ 0 & 0 & 2\mu \end{bmatrix} g.
    \label{SlowManifoldLifted}
\end{align}
When (\ref{SlowManifold}) has linear control $u\in\mathbb{R}^2$ and disturbance $d\in\mathbb{R}^2$,
\begin{align}
    \dot{x} =  
    \begin{bmatrix} \mu x_1 \\ \lambda (x_2 - x_1^2) \end{bmatrix} + u + d,
    \label{SlowManifoldControl}
\end{align}
the same lift yields a nonlinear system \cite{brunton2016koopman, kaiser2021data}, 
\begin{align}
    \dot{g} = 
    \begin{bmatrix} \mu & 0 & 0 \\ 0 & \lambda & - \lambda \\ 0 & 0 & 2\mu \end{bmatrix} g + \begin{bmatrix} 1 & 0 \\ 0 & 1 \\ 2g_1 & 0 \end{bmatrix}(u + d).
    \label{SlowManifoldControlLifted}
\end{align}
We may fix $g_1$ in the control and disturbance matrix, but clearly this yields a local linearization which may be driven off the manifold of true lifted state, given by $g_1^2 = g_3$. 

\begin{figure*}[t]
    \centering
    \includegraphics[width=\linewidth]{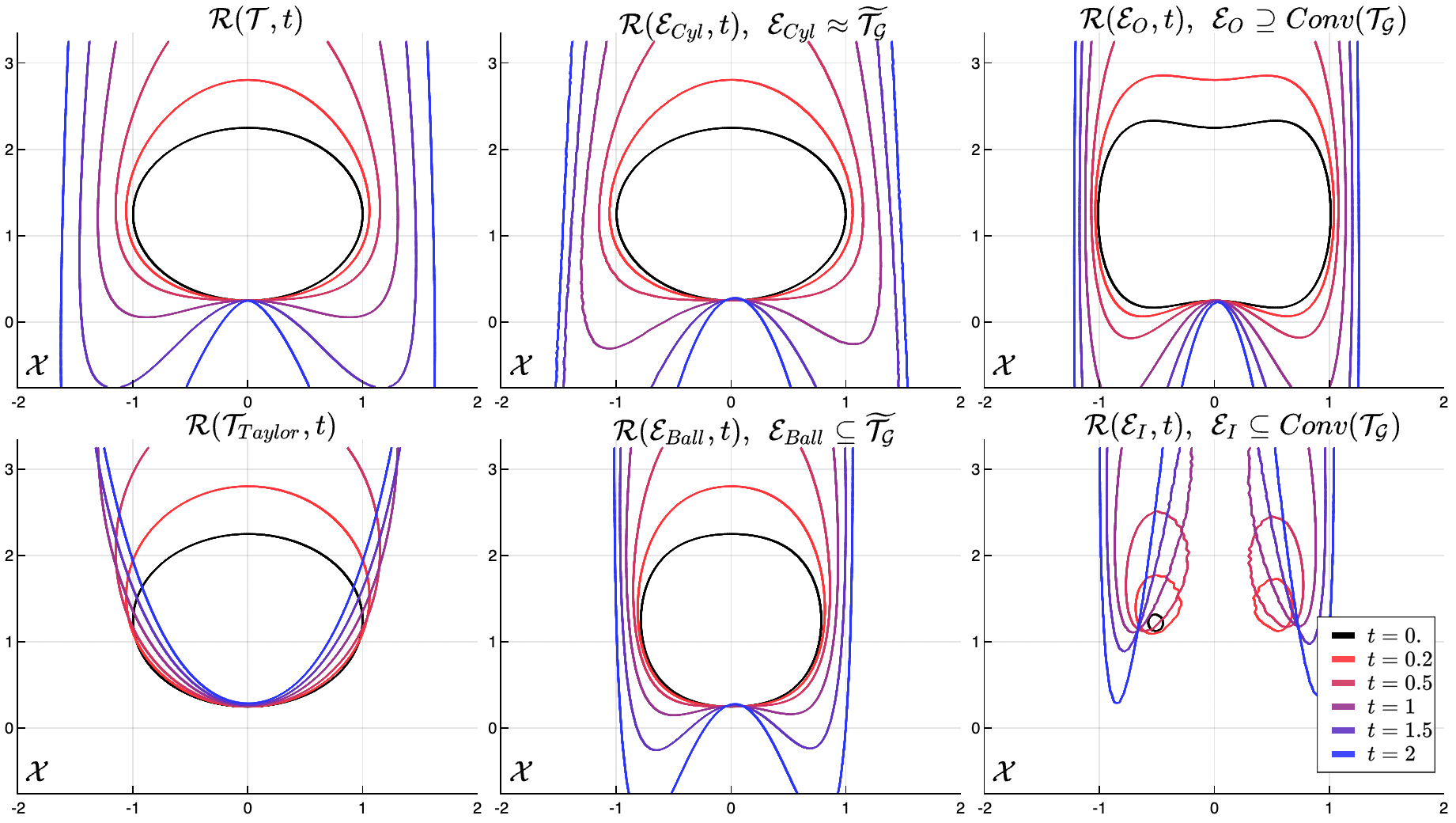} 
    \caption{\textbf{Koopman-hopf BRS Evolutions for Various Targets in the Slow Manifold System} The evolution of the BRS, interpolated as the zero-contour of the Hopf formula over a grid, is plotted for four lifted targets against the DP solution, $\mathcal{R}(\mathcal{T},t)$ (upper left), and local-linearization Hopf solution, $\mathcal{R}(\mathcal{T}_{Taylor},t)$ (bottom left), all of which are solved for $t=[0.2, 0.5, 1., 1.5, 2.]$ seconds (spanning red to blue, with the target, i.e $t=0$, in black). This is done for a convex control and disturbance game in the slow manifold system for which the autonomous dynamics are exactly linearizeable, but the control and disturbance dynamics are not.}
    % \caption{\textbf{BRS Evolutions and Quatitative Comparison for Various Targets in the Slow Manifold System} Top, the evolution of the BRS, interpolated as the zero-contour of the Hopf formula over a grid, is plotted for four defined elliptical targets against the DP-HJR solution, $\mathcal{R}(\mathcal{T},t)$, and Taylor-Hopf solution $\mathcal{R}(\mathcal{T}_{Taylor},t)$ solved for $t=[0.2, 0.5, 1., 1.5, 2.]$ seconds (spanning red to blue, target in black). This is done for a convex control and disturbance game in the slow manifold system for which the autonomous dynamics are exactly linearizeable, but the control and disturbance dynamics are not. Below, four metrics of the evolutions are displayed for the four targets: 1. the Jaccard Index for comparing sets $JI$ (bottom, upper-left), 2. the run-time per point (bottom, bottom-left) 3. the percentage (over the grid) of Falsely-Included points $FI\%$ (bottom, upper-right) 4. the percentage of Falsely-Excluded $FE\%$ (bottom, upper-right). }
    \label{fig:slowmanifold}\vspace{-1em}
\end{figure*}

\begin{figure}
    \centering
    \includegraphics[width=0.75\linewidth]{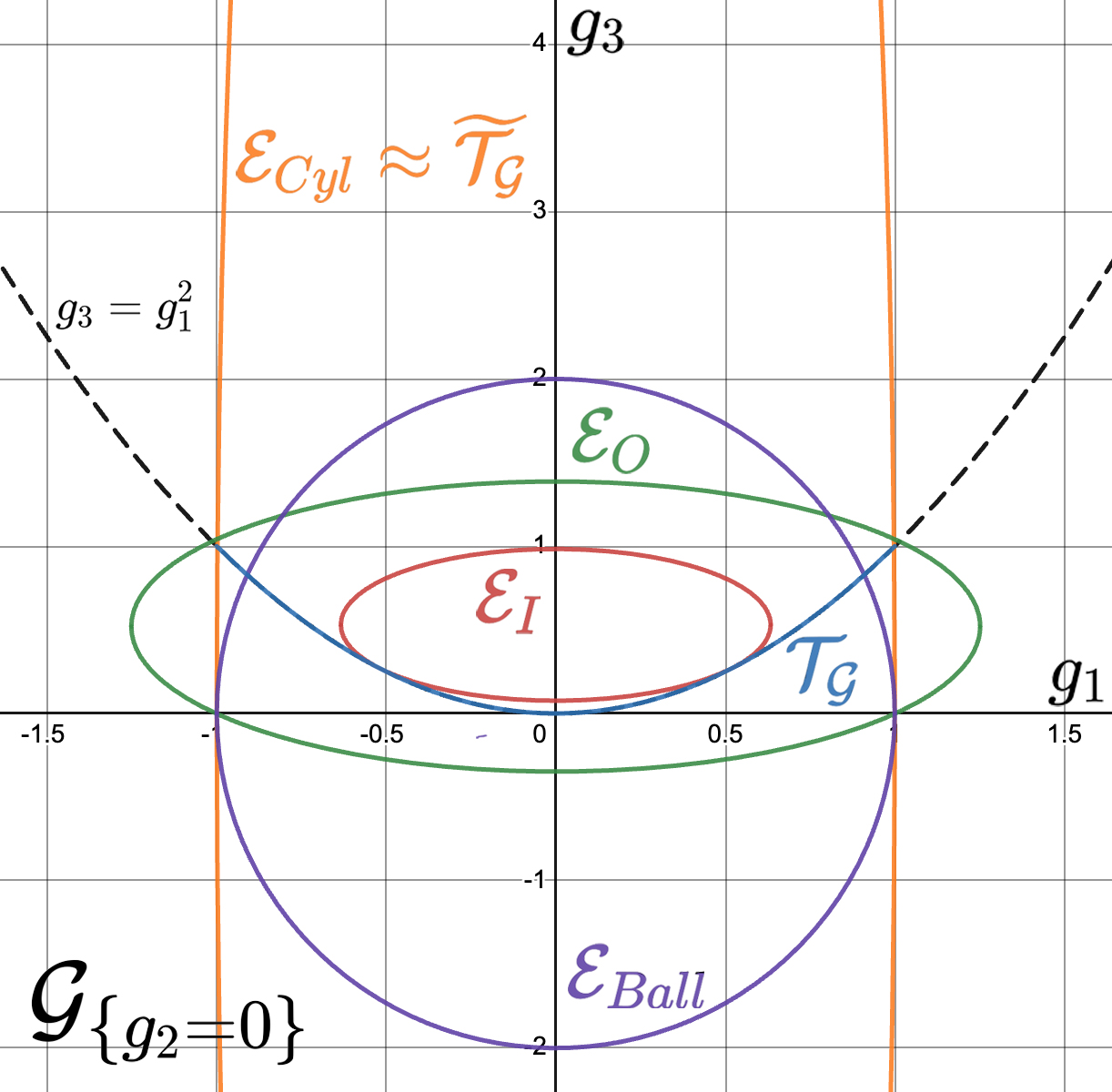}    
    \caption{\textbf{Lifted Targets for the Slow Manifold} This illustration depicts the four targets considered in the the $g_2 = 0$ slice of the slow manifold Koopman space $\mathcal{G}$: the prolate ellipse approximately equal to the augmented lifted target, $\mathcal{E}_{Cyl} \approx \widetilde{\mathcal{T}}_\mathcal{G}$, the ball, $\mathcal{E}_{Ball} \subseteq \widetilde{\mathcal{T}}_\mathcal{G}$, the outer ellipse of the lifted target hull, $\mathcal{E}_{O} \supseteq Conv(\mathcal{T}_\mathcal{G})$, the inner ellipse of the lifted target hull, $\mathcal{E}_{I} \subseteq Conv(\mathcal{T}_\mathcal{G})$. }
    \label{fig:LiftedTargets}\vspace{-1em}
\end{figure}

One might question the use of Koopman theory for this application since there exists a local linearization, however, the success of this approach (particularly when coupled with MPC) has significantly out-performed other linearization methods \cite{korda2018linear, kaiser2021data, bruder2019modeling}. This motivates the use of these approximate Koopman forms to accommodate trajectories that evolve off the manifold of lifted true states. In the proposed work, omitting the end of these trajectories from the lifted target might omit their origin on the manifold in the backwards-reachable expansion. Hence, we consider four targets in the lifted space and compare their Hopf-solved, BRS evolution for a convex game between control and disturbance. 

Consider a reach problem in (\ref{SlowManifoldControl}) with a target $\mathcal{T}$ given by a ball of radius $r^x:=1$ centered at $c^x:=[0, 1.25]$ and control and disturbance bounded by balls at the origin with radii of $r^u:=0.5$ and $r^d:=0.25$ respectively. We use the approximate Koopman formulation given in (\ref{SlowManifoldControlLifted}), defined by lifting function $\Psi_{SM}$, with $g_1:=0$ fixed in the input matrix. Following Sec. \ref{sec:koopman_hopf}, with the image of the true target under the lifting function, $\mathcal{T}_\mathcal{G}$, and pre-image under the lowering function, $\widetilde{\mathcal{T}}_\mathcal{G}$, we may consider several elliptical targets in the lifted space $\mathcal{G}$. A 2D slice of these lifted targets and corresponding reachable sets are visualized in Figure~\ref{fig:LiftedTargets}.
First, we define two targets in the Koopman space based on relaxations (\ref{PolynomialTargetApprox}) of the augmented lifted target $\widetilde{\mathcal{T}}_\mathcal{G}$: $\mathcal{E}_{Ball}:=\mathcal{C}_\mathcal{E}(I)$, a ball, and $\mathcal{E}_{Cyl}:=\mathcal{C}_\mathcal{E}(\text{Diag}[1,1,\eps])$, a prolate ellipse with $ \eps:=15$. Note, the third ellipse axes of these targets, along the lifted state $g_3$, are of value greater than zero, hence, $\mathcal{E}_{Ball} \subseteq \widetilde{\mathcal{T}}_\mathcal{G}$ and $\mathcal{E}_{Cyl} \subseteq \widetilde{\mathcal{T}}_\mathcal{G}$, and in the given grid $\mathcal{E}_{Cyl} \approx \widetilde{\mathcal{T}}_\mathcal{G}$. Second, we define $\mathcal{E}_O$ and $\mathcal{E}_I$ to be the outer and inner ellipses of the convex hull of the lifted target, $Conv(\mathcal{T}_\mathcal{G})$, solved by semi-definite programs \cite{ben2001lectures}. By definition, these are the unique, minimum-volume enclosing and unique, maximum-volume enclosed ellipses respectively and, hence, $\mathcal{E}_O \supseteq Conv(\mathcal{T}_\mathcal{G})$ and $\mathcal{E}_I \subseteq Conv(\mathcal{T}_\mathcal{G})$.  

With the procedure Algorithm~\ref{alg:KHP}, we plot the BRS evolution for each lifted target on a $100 \times 100$ grid in $\mathcal{X}$ lifted to $\mathcal{G}$. In Figure~\ref{fig:slowmanifold}, the top left plot shows the standard  HJR DP method using \textit{hj\_reachability.py} \cite{StanfordASL_hj_reachability}, which is treated as the ground truth. 
The bottom left plot shows the Hopf formula solved with the Taylor (rather than Koopman) linearization of (\ref{SlowManifoldControl}) with the original target $\mathcal{T}$, called $\mathcal{T}_{Taylor}$ for clarity, linearized at $c^x$. All other plots show our Koopman-Hopf method with the various target sets (outlined in black). The BRSs at $t = [0., 0.2, 0.5, 1, 1.5, 2]$ look-back times ($T=0$) are plotted for each lifted target. The contours are solved with the marching squares method using the centripetal Catmull-Rom interpolation scheme \cite{plotly, barry1988recursive}.

For a quantitative comparison of the results, we compare the $\mathcal{R}(\cdot,t)$ of each result to the DP solution $\mathcal{R}(\mathcal{T},t)$ with three metrics. The set similarity is quantified by the Jaccard Index \cite{gilbert1884finley}, 
\beqn
JI(\mathcal{R}_1,\mathcal{R}_2)= \frac{|\mathcal{R}_1 \cap \mathcal{R}_2|}{|\mathcal{R}_1 \cup \mathcal{R}_2|}.
\label{Jaccard}
\eqn
 over the shared grid ($1 \implies$ perfect matching). We also quantify the percentage of False Included and Excluded,
% \beqn
% FI\%(\mathcal{R}_1,\mathcal{R}_2)= \frac{|\{x \in \mathcal{R}_1 \land x \notin \mathcal{R}_2\}|}{|grid|}, \\
% FE\%(\mathcal{R}_1,\mathcal{R}_2)= \frac{|\{x \notin \mathcal{R}_1 \land x \in \mathcal{R}_2\}|}{|grid|}.
% \label{FIFE}
% \eqn
\beqn
\begin{aligned}
FI\%(\mathcal{R}_1,\mathcal{R}_2)= |\mathcal{R}_1 \setminus (\mathcal{R}_1 \cap \mathcal{R}_2)| \:/  \: |grid|, \\
FE\%(\mathcal{R}_1,\mathcal{R}_2)= |\mathcal{R}_2 \setminus (\mathcal{R}_1 \cap \mathcal{R}_2)| \: /  \:|grid|.
\end{aligned}
\label{FIFE}
\eqn
These measures, along with the average run time per point for each of the approximations, are averaged over $t=[0.2, 0.5, 1.0, 1.5, 2.0]$ and tabulated in Table~\ref{table:BRSresults_SM}.

\begin{table}[h]
\caption{Mean BRS similarity metrics (best-scoring is bolded) for each of the four lifted targets and the Taylor solution $\mathcal{R}(\mathcal{T}_{Taylor})$ compared to true DP solution, $\mathcal{R}(\mathcal{T})$. Note, the $t$ argument of $\mathcal{R}$ has been dropped to indicate that the results have been averaged over the look-back times $t=[0.2, 0.5, 1.0, 1.5, 2.0]$. The mean comp. time per-point is also given as $t_c$ in milliseconds $(ms)$. Note that each point is computed in parallel.} 
\label{table:BRSresults_SM}
\begin{center}
\renewcommand{\arraystretch}{1.3}
\begin{tabular}{|c||c|c|c|c|c|}
\hline
 & $\mathcal{R}(\mathcal{T}_{Taylor})$ & $\mathcal{R}(\mathcal{E}_{Cyl})$ &  $\mathcal{R}(\mathcal{E}_{Ball})$ & $\mathcal{R}(\mathcal{E}_{O})$ & $\mathcal{R}(\mathcal{E}_{I})$  \\
\hline
\hline
  $JI$ & 0.71  & \textbf{0.94} & 0.69  & 0.79 & 0.17   \\
  % \hline
 $FI\%$ & 0.82 & 1.18 & 0.04 & 4.64 & \textbf{0.00}    \\
 % \hline
 $FE\%$ & 17.1 & \textbf{2.3} & 16.6 & 7.3 & 40.7    \\
 % \hline
 $t_c$ & \textbf{3.3} & 33.0 & 5.6 & 5.7 & 18.5    \\
\hline
\end{tabular}
\end{center}
\vspace{-2mm}
\end{table}

The BRS plots and Jaccard data demonstrate the best approximation of the true reachable evolution is that of $\mathcal{E}_{Cyl}$, approximately equal to the augmented lifted target. Thus, many trajectories in this approximate Koopman form beginning at lifted true states must be driven off the manifold under the optimal control and disturbance in a manner akin to true evolution. The drawback of using the cylindrical ellipse is the significant increase in run-time per point, due to parameter $\eps$ determining the gradient of the optimization objective, the Hopf formula (\ref{HopfFormulaZ}). If speed were the utmost priority and processors limited, a complete BRS analysis might require solving the simpler lifted target $\mathcal{E}_{Ball}$ first to warm-start the solution of $\mathcal{E}_{Cyl}$. 

The inner and outer ellipses may be applied when seeking to minimize the percentage of falsely included or excluded points, respectively. We note, however, these to no provide guaranteed under or over approximations of the true BRS due to the approximate nature of the Koopman dynamics and further work is required to give analytic error bounds. 

In summary, we see that the choice of lifted target has a significant implication on the BRS approximation, even for a problem with exactly linearizeable autonomous dynamics due to the inclusion of inputs. Nonetheless, the results here demonstrate that the proposed method is viable for approximating the BRS evolution in a manner parallelizable over space which does not scale exponentially with dimension.

% big version
% \begin{figure*}
%     \centering
%     \includegraphics[height=0.9\textheight, keepaspectratio]{BRS_pannel_c_nc.png}    
%     \caption{\textbf{BRS Evolutions for Convex and Non-Convex Games in the Duffing System} Here the $\pm \epsilon$-boundary of a target $\mathcal{T}$ (black), the DP-solved BRS $\mathcal{R}(\mathcal{T},t)$ (blue), the local-linearization Hopf BRS $\mathcal{R}( \mathcal{T}_{\text{Taylor}},t)$ (green), and our 15D Koopman-Hopf BRS $\mathcal{R}(\widetilde{\mathcal{T}}_\mathcal{G},t)$ (gold) are plotted (in scatter to emphasize the space-parallelization of Hopf). In the left column, these correspond to a control \& disturbance convex game, in which disturbance only perturbs controlled states, and in the right column, to a non-convex game, in which disturbance perturbs all states. Both games are based in the Duffing oscillator and the BRS are plotted for three values of $t$ with $\epsilon=0.1$. With uniform optimizer parameters, higher error is observed at higher $t$, coinciding with the larger integral value in the Hopf objective (\ref{HopfFormula}). Additionally in the non-convex case, error arises from both Koopman lifting but also the discrepancy between minimax-viscosity and viscosity solutions (\cite{rublev2000generalized, subbotin1996minimax, chow2019algorithm}).}
%     \label{fig:BRS_pannel}\vspace{-1em}
% \end{figure*}

\subsection{Approximate BRS of the Duffing Oscillator}\label{subsec:BRSapprox}

Here, we test the Koopman-Hopf approach on the nonlinear Duffing oscillator given by the following dynamics,
\begin{align}
    \begin{bmatrix} \dot{x}_1 \\ \dot{x}_2 \end{bmatrix} =  
    \begin{bmatrix} x_2 \\ \alpha x_1 - \beta x_1^3 - \delta x_2 \end{bmatrix} + 
    \begin{bmatrix} 0 \\ 1 \end{bmatrix} u
    + B_2 d
    \label{DuffingSystem}
\end{align}
with $\alpha, \beta, \delta = 1, 1, 0.1$. We consider two problems: the system with disturbance only on the controlled states ($B_2 = B_1$), yielding a convex Hamiltonian, and the system with disturbance on the full state ($B_2 = I$), yielding a non-convex Hamiltonian (see after (\ref{InputConstraint})). In both cases, we consider a target $\mathcal{T}$ defined by a ball with radius of $r=2$. 

To generate an accurate Koopman approximation, we chose to use the 4-degree polynomial $P_4(x)$ (with 15 dimensions) as the lifting-function and fit the Koopman matrices $K$, $L_1$ and $L_2$ with the hyper-tuning survey performed by \textit{autokoopman.py} package \cite{Bak, EthanJamesLew_AutoKoopman}. Given this choice of lifting function containing the identity mapping, the definition of the augmented lifted target $\widetilde{\mathcal{T}}_\mathcal{G}$ is simply an infinite cylinder with radius $r=2$, which we relaxed to a prolate ellipse with $\eps=10$ in the non-identity states of the lift. We solve the BRS evolving from the augmented lifted target with the procedure outlined in Algorithm~\ref{alg:KHP}.

\begin{figure}
    \centering
    \includegraphics[width=\linewidth]{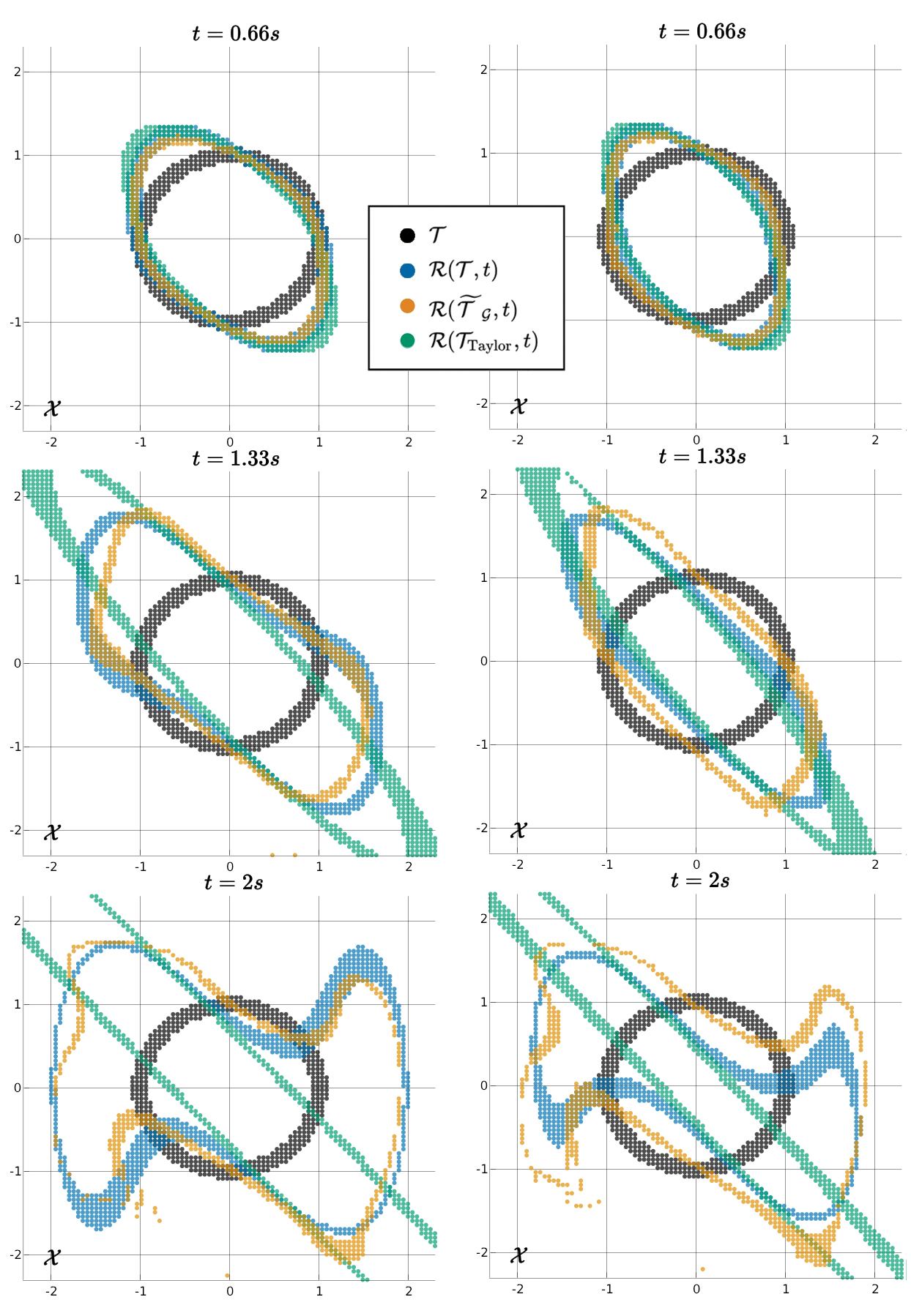}    
    \caption{\textbf{Koopman-Hopf BRS Evolutions for Convex and Non-Convex Games in the Duffing System} Here the $\pm \epsilon$-boundary ($\epsilon=0.1$) of a target $\mathcal{T}$ (black), the DP-solved BRS $\mathcal{R}(\mathcal{T},t)$ (blue), the local-linearization Hopf BRS $\mathcal{R}( \mathcal{T}_{\text{Taylor}},t)$ (green), and our 15D Koopman-Hopf BRS $\mathcal{R}(\widetilde{\mathcal{T}}_\mathcal{G},t)$ (gold) are plotted. In the left column, these correspond to a convex control \& disturbance game (disturbance only perturbs controlled states), and in the right column, to a non-convex game (disturbance perturbs all states). Both games are based in the Duffing oscillator and the BRS are plotted for three values of $t$. 
    % With uniform optimizer parameters, higher error is observed at higher $t$, coinciding with the larger integral value in the Hopf objective (\ref{HopfFormula}). Additionally in the non-convex case, error arises from both Koopman lifting but also the discrepancy between minimax-viscosity and viscosity solutions (\cite{rublev2000generalized, subbotin1996minimax, chow2019algorithm}).}
    }
    \label{fig:BRS_pannel}\vspace{-1em}
\end{figure}

We compare our results with an HJR DP method, \textit{hj\_reachability.py} \cite{StanfordASL_hj_reachability} as well as the Hopf formula with the true target solved with the Taylor approximation of the system, called $\mathcal{T}_{\text{Taylor}}$. Figure~\ref{fig:BRS_pannel} compares the BRSs at $t = [0.66, 1.33, 2]$ look-back times ($T=0$) for both the convex problem (with disturbance on controlled states only and of less-than or equal magnitude), and non-convex problem (with disturbance on all states). 

We quantify the similarity of the sets with the Jaccard Index \cite{gilbert1884finley} over a common discritized grid. We additionally include a baseline derived from the Taylor series approximate dynamics with a localization point at the center of the target in $\mathcal{X}$. The numerical results can be viewed in Table~\ref{table:BRSresults}.

% small version of pannel

\begin{table}[h]
\caption{Approximate BRS Jaccard $JI$ similarity to the DP solution, $\mathcal{R}(\mathcal{T}, \cdot)$ for each time in the disturbance on control problem i.e convex game (left), and disturbance on all states problem i.e. non-convex game (right). Best-scoring is bolded.}
\label{table:BRSresults}
\renewcommand{\arraystretch}{1.3}
\begin{center}
% \begin{tabular}{c|c c|c c|}
%  & \multicolumn{2}{c|}{\textbf{Convex Game}} & \multicolumn{2}{c}{\textbf{Non-Convex Game}} \\
% \end{tabular}
\begin{tabular}{|c|c|c||c|c|}
% \cline{2-5}
\hline
 & \multicolumn{2}{c||}{\textbf{Convex Game}} & \multicolumn{2}{c|}{\textbf{Non-Convex Game}} \\
% \cline{2-5}
% \hline
% \hline
% \cline{0-0}
t & $\mathcal{R}(\mathcal{T}_{Taylor}, \cdot)$ & $\mathcal{R}(\widetilde{\mathcal{T}}_\mathcal{G}, \cdot)$ &  $\mathcal{R}(\mathcal{T}_{Taylor}, \cdot)$ & $\mathcal{R}(\widetilde{\mathcal{T}}_\mathcal{G}, \cdot)$  \\
\hline
\hline
  0.0 & \textbf{1.0}  & 0.97 & \textbf{1.0}  & 0.97   \\
 0.33 & 0.90 & \textbf{0.96} & \textbf{0.92} & 0.91    \\
 0.66 & 0.80 & \textbf{0.88} & \textbf{0.85} & 0.80    \\
 0.99 & 0.61 & \textbf{0.82} & \textbf{0.71} & 0.66    \\
 1.32 & 0.47 & \textbf{0.78} & 0.50 & \textbf{0.60}    \\
 1.65 & 0.37 & \textbf{0.79} & 0.37 & \textbf{0.59}    \\
 1.98 & 0.30 & \textbf{0.76} & 0.30 & \textbf{0.58}    \\
\hline
\end{tabular}
\end{center}
\vspace{-3mm}
\end{table}

In Figure~\ref{fig:BRS_pannel} and Table~\ref{table:BRSresults}, many important results are apparent. As in the slow manifold system, the Koopman-Hopf method has error that arises immediately from definition of the inexact lifted target, which is not identical to $\mathcal{T}_\mathcal{G}$. Additionally, in the non-convex problem we lose guaranteed convergence to the global optimum of the Hopf formula (the minimax solution) and, separately, agreement between the minimax and viscosity solutions \cite{rublev2000generalized, subbotin1996minimax, chow2019algorithm}, causing further discrepancy between the DP solution and the Taylor and Koopman based Hopf methods in the right columns. In both cases its possible to observe that the Taylor series estimate becomes poor for long horizons while the Koopman-Hopf solution is more robust to error. Particularly in the convex case, its possible to observe that the the Taylor method falls to having only a $30\%$ agreement (by Jaccard), while the Koopman-Hopf method remains above $75\%$ up to $t=2s$. With improved Koopman and Hopf methods, we expect these results to extend to the analysis and control of more complex, high-dimensional nonlinear systems.

\subsection{Comparison of Koopman control in the Glycolysis Model}\label{subsec:Glycolysis}

\begin{figure*}
    \centering
    \includegraphics[width=\linewidth]{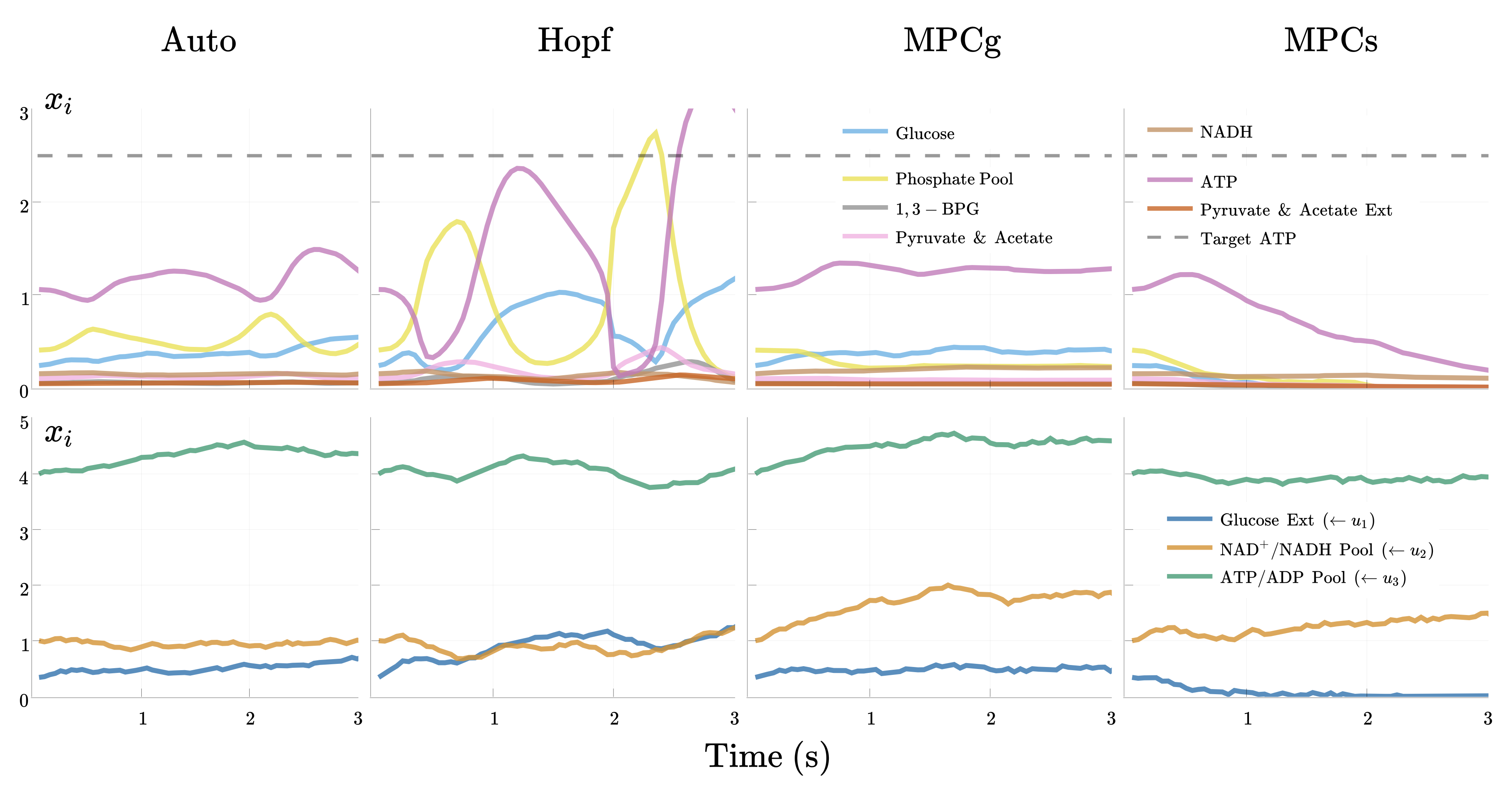}    
    \caption{\textbf{Comparison of Koopman-based Controllers in a Glycolysis System} Three controlled evolutions, using the same Koopman lift, of the 10D glycolysis model with the same disturbance trajectory and initial condition. ``Auto'' signifies the disturbed autonomous system. The bottom pannels plot the three controlled states in the model that can be controlled. Koopman-Hopf controller (``Hopf" above) amplifies the cycle of the phosphate, glucose and ATP states by oscillating the total concentrations of NAD+/NADH and ATP/ADP to achieve the target in a manner that translates to the nonlinear system.}
    \label{fig:glycolysis}\vspace{-1em}
\end{figure*}

Next, we compare the ability of a Koopman-Hopf controller to navigate a 10D glycolysis model \cite{yeung2019learning,ruoff2003temperature,daniels2015efficient},
\beqn
\begin{aligned}
\dot{x}_1 & =\kappa (x_1 - x_7) -\frac{k_1 x_1 x_6}{1+\left(x_6 / K_1\right)^4} \\
\dot{x}_2 & =2 \frac{k_1 x_1 x_6}{1+\left(x_6 / K_1\right)^4}-k_2 x_2 ( x_9 - x_5 )  -k_6 x_2 x_5 \\
\dot{x}_3 & =k_2 x_2 (x_9 - x_5)  - k_3 x_3 ( x_{10} - x_6 ) \\
\dot{x}_4 & =k_3 x_3 (x_{10} - x_6) - k_4 x_4 x_5 - \kappa (x_4 - x_8) \\
\dot{x}_5 & =k_2 x_2 (x_9 - x_5)  - k_4 x_4 x_5 - k_6 x_2 x_5 \\
\dot{x}_6 & =-2 \frac{k_1 x_1 x_6}{1+\left(x_6 / K_1\right)^4}+2 k_3 x_3 (x_{10} - x_6)-k_5 x_6 \\
\dot{x}_7 & = u_1 + d_1 \\
\dot{x}_8 & =\mu \kappa (x_4 - x_8) - k x_8  \\
\dot{x}_9 & = u_2 + d_2 \\
\dot{x}_{10} & = u_3 + d_3,
\end{aligned}
\label{glycolysis}
\eqn
where the components of $x$ are $[$glucose, phosphate pool, 13-BPG, pyruvate/acetaldehyde pool, NADH, ATP, extracellular glucose, extracellular pyruvate/acetaldehyde, NAD+/NADH pool, ATP/ADP pool$]$ and the values of all parameters are inherited from \cite{yeung2019learning}. The model captures the nonlinear enzymatic reaction network for conversion of glucose with cellular currencies ATP and NADH. Note, the dimension of the system makes DP-based HJ optimal control infeasible.

We use the trivial Koopman form corresponding to a 10-dimensional DMDc linearization, which we compute with the \textit{pykoopman.py} package \cite{dynamicslab_pykoopman}. The Koopman-Hopf controller uses the standard Hopf controller formulation  \cite{donggun19iterativehopf,doggn2020hopf,Kirchner_2018} to solve for the minimum time $T^*$ when the target is reachable for the current state $x$,
\beqn
T^* = \text{argmin}_T \phi(x, t) = \text{argmin}_T \phi(g, t), \quad g = \Psi(x).
\label{MinT}
\eqn
The controller then applies the optimal control at this time which is derived from,
\beqn
\nabla_p H(\nabla_{g_\mathcal{Z}} \phi(g_\mathcal{Z}, T^*), T^*) = e^{-tK} L_1 u^* + e^{-tK} L_2 d^*.
\label{HJocHopf}
\eqn

We compare this Koopman-Hopf controller with two Koopman-MPC formulations that incorporate the same Koopman system \cite{bemporad2002model, korda2018linear}. The game MPC (MPCg) solves the optimal control for a random fixed disturbance with a one-step horizon, the disturbance then does the same for the previous control, and the process repeats until improvment plateaus. The stochastic MPC (MPCs) \cite{mesbah2016stochastic} samples 20 random disturbance trajectories and minimizes the expectation of the cost given evolution with those samples. In both cases, the MPC algorithms are defined with a horizon of 10 steps of 0.1 seconds (where performance plateaued), and are defined by a terminal cost alone (no running cost) to match the Hopf controller. 

All controllers (Hopf and MPC) are based on the same terminal value $J$ which implicitly only defines a target ATP concentration. This encourages the controllers to achieve a specific concentration of ATP, as a bioengineer might desire for cell growth, by manipulating the external glucose, total NAD+/NADH pool, and ADP/ATP pool. This was implemented with another prolate ellipse which is tight only on the ATP dimension, and because of the trivial nature of the lift this target is identical to its lifted forms (\ref{IdentityTarget}). We chose input sets $\mathcal{U}$ and $\mathcal{D}$ to be ellipses (\ref{InputConstraint}) to couple the controls, as common in biological problems, and assume disturbance on actuation only (convex Hamiltonian).

After each controller computes an estimate optimal action $\hat u^*(t)$, the system is then progressed with the true nonlinear dynamics (computed with Radau \cite{hairer1999stiff}) and a disturbance. 

The simulation was run 50 times, with each controller subjected to the same random initial conditions (sampled from the realistic concentration bounds given in \cite{ruoff2003temperature, daniels2015efficient}) and the same random disturbance trajectory. Results are shown in Table~\ref{MPCresults} and Figure~\ref{fig:glycolysis}. 

\begin{table}[h]
\caption{Controller Results for 50 random $x_0$ bounded by \cite{daniels2015efficient} and random $d(\cdot)$. The mean comp. time per-point is given as $t_c$. The best-scoring is bolded.}
\label{MPCresults}
\begin{center}
\renewcommand{\arraystretch}{1.3}
\begin{tabular}{|c||c|c|c|c|}
\hline
Controller & Success \% & Mean, Max ATP (m) & $t_c$ (s) \\
\hline
\hline
Auto & 20 & 2.07 & - \\
Hopf & \textbf{96} & \textbf{3.06} & 0.81 \\
MPCg & 22 & 2.14 & \textbf{0.02} \\
MPCs & 16 & 1.96 & 0.36 \\
\hline
\end{tabular}
\end{center}
\vspace{-4mm}
\end{table}

The system is complex because of the highly-stiff nonlinearities, state constraints ($x_i > 0$) and inter-connectivity of the metabolic network; naively driving the ATP/ADP pool up leads to counter-productive results, as demonstrated by MPCg and MPCs in Figure~\ref{fig:glycolysis}.

The Koopman-Hopf controller appears to overcome the nonlinearity of the system by amplifying the oscillations of ATP to reach the target. From these results, and particularly the failure of the MPCs, navigation to this goal appears to require some long-horizon planning which may be based on a medium-accuracy (linear) model that lacks gratuitous high-resolution fidelity. 
% We might think of this as a ``smoothing'' of the nonlinear dynamics, such that the major trends of flows are captured, but in general, lack gratuitous, high-resolution fidelity. 
Furthermore, the observed success is due to the Koopman-Hopf controller's robustness to disturbance which materialized as model perturbation but also error in the lift. 
Note, we have no guarantees on the success despite starting in the BRS in the Koopman model. It is also possible be driven to the border of state constraints, despite their implicit effect on the Koopman model generation; to avoid this one would need to cautiously consider the Hopf formula with state constraints \cite{lee2023efficient}. Nonetheless, it is clear from the results that the Koopman-Hopf controller often succeeds despite true disturbance and state constraints. It does this by efficiently computing robust trajectories, unlike the MPC, making the proposed framework a worthy approach for high-dimensional, long-horizon planning.

\section{Conclusion}\label{sec:conclusion}

We propose a novel Koopman-Hopf method in order to approximate Hamilton-Jacobi reachability in high-dimensional, nonlinear systems. We find this approach works well for approximating BRSs and driving high-dimensional, nonlinear systems with bounded disturbance. We hope to extend this work along several exciting directions, including expansion to more complicated lifting functions such as radial basis functions and neural networks, applying this to black box systems, and quantifying the uncertainty based on the Koopman linearization error \cite{haseli2022temporal}. 
Moreover, with more advanced methods for Hopf optimization, one might allow a broader class of reachability problems and Koopman methods and we leave this to future work.
% We believe that the proposed method is valuable for robustly maneuvering a range of otherwise intractable systems.

\section*{Acknowledgements}

We thank Drs. Steven Brunton, Stanley Bak, Ian Abraham and Masih Haseli for discussions about Koopman theory, Drs. Gary Hewer and Matthew Kirchner for discussions about applying Hopf to Naval applications, and Dr. Somil Bansal for thought-provoking feedback. Finally, we thank Zheng Gong and Sander Tonkens for valuable feedback.

\bibliographystyle{IEEEtran}
\bibliography{main}
\end{document}